\documentclass[aps,twocolumn,nofootinbib,showpacs,final,balancelastpage,superscriptaddress]{revtex4}
\usepackage{graphicx}
\usepackage{tabularx}
\usepackage{amssymb}
\usepackage{amsmath}
\usepackage{amsbsy}
\usepackage{comment}
\usepackage{mathrsfs}
\usepackage{dsfont}
\usepackage{nicefrac}
\usepackage{wrapfig}
\usepackage{amsfonts}
\usepackage{url} % link package
\usepackage[nolist]{acronym}
\usepackage{nicefrac}
\usepackage{subfigure}
\usepackage{epsf,color,colordvi,pifont}
\usepackage[dvipsnames]{xcolor}
\usepackage{showkeys}
\def\deltabar{{\mathchar'26\mkern-8mu\delta}}
\def\dbar{{\mathchar'26\mkern-12mu d}}
\DeclareFontFamily{OT1}{pzc}{}
\DeclareFontShape{OT1}{pzc}{m}{it}{<-> s * [1.10] pzcmi7t}{}
\DeclareMathAlphabet{\mathpzc}{OT1}{pzc}{m}{it}
\def\dbar{{\mathchar'26\mkern-12mu d}}

\begin{document}
\title{Axion-modified photon propagator,  Coulomb potential and  Lamb-shift}
\author{S. \surname{Villalba-Ch\'avez}}
\email{selym@tp1.uni-duesseldorf.de}
\affiliation{Institut f\"{u}r Theoretische Physik I, Heinrich-Heine-Universit\"{a}t D\"{u}sseldorf, Universit\"{a}tsstra\ss{e} 1, 40225 D\"{u}sseldorf, Germany.}
\author{A. \surname{Golub}}
\email{Alina.Golub@uni-duesseldorf.de }
\affiliation{Institut f\"{u}r Theoretische Physik I, Heinrich-Heine-Universit\"{a}t D\"{u}sseldorf, Universit\"{a}tsstra\ss{e} 1, 40225 D\"{u}sseldorf, Germany.}
%\author{A. \surname{Shabad}}
%\email{shabad@lpi.ru}
%\affiliation{P. N. Lebedev Physics Institute, Moscow 117924, Russia.}
\author{C. \surname{M\"{u}ller}}
\email{c.mueller@tp1.uni-duesseldorf.de}
\affiliation{Institut f\"{u}r Theoretische Physik I, Heinrich-Heine-Universit\"{a}t D\"{u}sseldorf, Universit\"{a}tsstra\ss{e} 1, 40225 D\"{u}sseldorf, Germany.}

\begin{abstract}
A consistent renormalization of a quantum theory of axion-electrodynamics requires terms beyond the minimal coupling of two photons to a neutral pseudoscalar field. 
This procedure is used to determine the self-energy operators of the electromagnetic and the axion fields with an accuracy of second-order in the axion-diphoton 
coupling. The resulting polarization tensor is utilized for establishing the axion-modified Coulomb potential of a static pointlike charge. In connection, the plausible 
distortion of the Lamb-shift in hydrogenlike atoms is established and the scopes for searching axionlike particles in high-precision atomic spectroscopy and in experiments 
of Cavendish-type are investigated. Particularly, we show that these hypothetical degrees of freedom are ruled out as plausible candidates for explaining the proton radius 
anomaly in muonic hydrogen. A certain loophole remains, though, which is linked to the nonrenormalizable nature of axion-electrodynamics. 
\end{abstract}



\date{\today}

\maketitle

%%%%%%%%%%%%%%%%%%%%%%%%%%%%%%%%%%%%%%%%%%%%%%%%%%%%%%%%%%%%%%%%%%%%%%%%%%%%%%%%%%%%%%%%%%%%%%%%%%%%%%%%%%%%%%%%%%%%%%%%%%%%%%%%%%%%
\section{Introduction}
%%%%%%%%%%%%%%%%%%%%%%%%%%%%%%%%%%%%%%%%%%%%%%%%%%%%%%%%%%%%%%%%%%%%%%%%%%%%%%%%%%%%%%%%%%%%%%%%%%%%%%%%%%%%%%%%%%%%%%%%%%%%%%%%%%%%

That the path integral measure in  quantum chromodynamics (QCD) is not invariant under an axial chiral $\rm U(1)_{A}-$transformation 
provides  a clear evidence  that this classical symmetry does not survive the quantization procedure.  As a consequence of this anomaly, 
QCD should not be a charge-parity ($CP$)-preserving framework. However, with astonishing experimental accuracy, no $CP$-violation event is 
known within the theory of the strong interactions. This so-called ``strong $CP$-problem'' finds a consistent theoretical solution by 
postulating a global $\mathrm{U(1)_{PQ}}$-invariance in the standard model (SM), which compensates the $CP$-violating term via its spontaneous 
symmetry breaking \cite{Peccei:1977hh}.  While this mechanism  seems to be  the most simple and robust among other possible routes of 
explanation, it is accompanied by a new puzzle linked to the nonobservation  of the associated  Nambu-Goldstone boson, i.e. the QCD axion 
\cite{Wilczek:1977pj,Weinberg:1977ma}. As a consequence, constraints resulting from this absence indicate a feeble interplay 
between this hypothetical particle and the well-established SM branch, rendering its detection a very challenging problem to 
overcome. Still, various experimental endeavours are currently oriented to detect this elusive degree of freedom  or, more generally, an 
associated class of particle  candidates  sharing its  main features, i.e. axionlike particles (ALPs). Some of them being central pieces 
in models which attempt to explain the dark matter abundance in our Universe  \cite{covi,Raffelt:2006rj,Duffy:2009ig,Sikivie:2009fv,Baer:2010wm}, 
whereas others are remnant features of string compactifications \cite{Witten:1984dg,Svrcek:2006yi,Lebedev:2009ag,LCicoli:2012sz}. 

The problematic associated with the ALPs detection demands both to exploit existing high-precision techniques and to develop new routes along   
which imprints of these hypothetical particles can be observed \cite{Jaeckel:2006id,evading}. Descriptions of the most popular detection methods can 
be found in Refs.~\cite{Jaeckel:2010ni,Ringwald:2012hr,Hewett:2012ns,Essig:2013lka,Alekhin:2015byh}. A vast majority of these  searches relies on the axion-diphoton 
coupling  encompassed within axion-electrodynamics  \cite{Wilczek:1987pj}. Correspondingly, many photon-related experiments such as those searching 
for light shining through a wall \cite{Chou:2007zzc,Afanasev:2008jt,Steffen:2009sc,Pugnat:2007nu,Robilliard:2007bq,Fouche:2008jk,Ehret:2010mh,Balou} 
and the ones based on polarimetry detections \cite{Cameron:1993mr,BMVreport,Chen:2006cd,Mei:2010aq,DellaValle:2013xs} have turned out to be particularly 
powerful. 

Contrary to that, precision tests of the Coulomb's law via atomic spectroscopy and experiments of Cavendish-type have not been used so far in the search 
for ALPs, although they are known to constitute  powerful probes for other well-motivated particle candidates \cite{Popov,Glueck,Jaeckel:2009dh,Roy}. In  
particular, these setups provide the best laboratory bounds on minicharged particles in the sub $\mu eV$ mass range. Simultaneously, by  investigating the 
role of ALPs in atomic spectra, one might elucidate whether the quantum vacuum of these hypothetical degrees of freedom may be the source for the large discrepancy  
$>5\sigma$ between the proton radius that follows from the Lamb-shift in muonic hydrogen versus the established value based on  electron scattering and  the Lamb-shift 
in ordinary hydrogen \cite{Pohl,antognini,Pohl2016,Beyer}. Various theoretical investigations have been put forward  seeking for a satisfactory explanation 
for this anomaly \cite{barger,brax,Jentschura:2014ila,Jentschura:2014yla,Liu}, some of them including hypothetical scalar particles. 

Against the background of these circumstances, it is relevant to derive modifications of the Coulomb potential due to quantum vacuum fluctuations of 
axionlike fields and to study their potential consequences. The former are encompassed  in the corresponding vacuum  polarization tensor whose calculation,  
however, is not a straightforward task  as far as  axion quantum electrodynamics  ($\rm QED_A$) is concerned. This is because 
it requires--first  of all--a  meaningful implementation of the corresponding perturbative expansion in this nonrenormalizable framework. 
In analogy to quantum gravity \cite{thooft,Stelle:1976gc,Goroff:1985sz,Goroff:1985th,Donoghue:1994dn,Donoghue:1993eb,Gies:2016con}, 
the expansion in terms of the axion-diphoton coupling gives rise to an infinite number of divergencies that cannot be reabsorbed in the renormalization 
constants associated with the parameters and fields of the theory. Unless a similar amount of counterterms is added, this feature spoils the predictivity 
of the corresponding scattering matrix,  preventing the construction of a consistent  quantum theory of axion-electrodynamics. Hence, the perturbative 
renormalizability of this theory demands unavoidably the incorporation of higher dimensional operators. This, in turn, comes along with the presence 
of a large number of free parameters which have to be fixed from experimental data. It is, however, known that this formal aspect relaxes because  many 
of these higher-dimensional terms are redundant, in that the ultimate scattering matrix is not sensitive to their coupling constants \cite{Arzt:1993gz,GrosseKnetter:1993td,Georgi:1991ch,Einhorn:2001kj}. 
As a matter of fact, all contributions of this nature can be formally eliminated while the effective Lagrangian acquires counterterms  which allow for 
the cancellation of the loop divergences. Besides,  when working at a certain level of accuracy, only a finite number of counterterms is needed and the 
cancellations of the involved infinities can be carried out pretty much in the same way as in conventional renormalizable field theories. 

In this paper, axion-electrodynamics is regarded  as a Wilsonian effective theory parametrizing the leading order contribution of an ultraviolet completion linked to 
physics beyond the SM. Its quantization is used for determining the self-energy operator of the electromagnetic field  with an accuracy of second-order 
in the axion-diphoton coupling. This result is utilized then  for obtaining the modified Coulomb potential of a static pointlike charge. In connection, the 
plausible distortion of the Lamb-shift in hydrogenlike atoms is established. Particular attention is paid to a limitation caused by the nonrenormalizable 
feature of axion electrodynamics which prevents us from having a precise and clear picture of the axion physics at distances smaller than the natural cutoff 
imposed by the axion-diphoton coupling. In contrast to previous studies of the Lamb-shift  involving minicharged particles and hidden photon fields, this property 
introduces an unknown uncertainty that cannot be determined, unless the ultraviolet completion  of axion-electrodynamics is found.  We argue that--up to 
this uncertainty--axionlike particles are ruled out as plausible candidates for explaining the proton radius anomaly in muonic hydrogen  ($\rm H_\mu$). Parallelly, 
spectroscopic results linked to a variety of transitions in hydrogen are exploited to probe the sensitivity of this precision technique in the search for ALPs. 
We show that, as a consequence of the mentioned feature, high-precision spectroscopy lacks of sufficient sensitivities as to improve the existing laboratory constraints 
on the parameter space of ALPs.

Our treatment is organized as follows.  Firstly, in Secs.~\ref{sec:QKAA} and \ref{poMS}, techniques known from effective field theories are exploited  for establishing  
the vacuum polarization tensor within an accuracy of the second order in the axion-diphoton coupling. There we show that, despite the electrically neutral nature 
of ALPs, the polarization tensor closely resembles the one obtained in QED. The similarity is stressed even further in Secs.~\ref{ACPseca} and \ref{ACPsecb}, where various 
asymptotes of the polarization tensor are established and the general expression for the  axion-Coulomb potential is determined. The latter outcome is presented in 
such a way that a direct  comparison with the Uehling potential can be carried out. Also in Sec.~\ref{ACPsecb}, we derive the corresponding modification to the 
Lamb-shift and emphasize the problematic introduced by both the effective scenario and the use of atomic  $s-$states. In Sec.~\ref{ACPsecc} we give some estimates and 
discuss the advantages and disadvantages of testing ALPs via excited states in  $\rm H_\mu$, whereas  in Sec.~\ref{conclus}  our conclusions are exposed. 
Some details about the  particle-ghost content of the theory are provided in Appendix~\ref{App1}. Finally,  in Appendix~\ref{Appendix}  the sensitivity levels associated 
with precision tests of the axion-modified Coulomb law via experiments of Cavendish-type are presented.

%%%%%%%%%%%%%%%%%%%%%%%%%%%%%%%%%%%%%%%%%%%%%%%%%%%%%%%%%%%%%%%%%%%%%%%%%%%%%%%%%%%%%%%%%%%%%%%%%%%
\section{The modified photon propagator in axion-electrodynamics} \label{sec:QKA}
%%%%%%%%%%%%%%%%%%%%%%%%%%%%%%%%%%%%%%%%%%%%%%%%%%%%%%%%%%%%%%%%%%%%%%%%%%%%%%%%%%%%%%%%%%%%%%%%%%%

%%%%%%%%%%%%%%%%%%%%%%%%%%%%%%%%%%%%%%%%%%%%%%%%%%%%%%%%%%%%%%%%%%%%%%%%%%%%%%%%%%%%%%%%%%%%%%%%%%%%%%%%%%%%%%%%%
\subsection{Effective field theory approach \label{sec:QKAA}}
%%%%%%%%%%%%%%%%%%%%%%%%%%%%%%%%%%%%%%%%%%%%%%%%%%%%%%%%%%%%%%%%%%%%%%%%%%%%%%%%%%%%%%%%%%%%%%%%%%%%%%%%%%%%%%%%%

Axion-electrodynamics relies on an effective action characterized by  a natural ultraviolet  scale $\Lambda_{\mathrm{UV}}$ at which the 
$\mathrm{U(1)_{PQ}}-$symmetry is broken spontaneously. It  combines the standard Maxwell Lagrangian, the free Lagrangian 
density of the  pseudoscalar field  $\bar{\phi}(x)$  and an interaction term coupling two photons and an axion. Explicitly,
\begin{equation}
\begin{split}
&S_{\bar{g}}= \int d^4 x \left\{- \frac{1}{4} f_{\mu \nu} f^{\mu \nu} + \frac{1}{2}\partial_{\mu}\bar{\phi}\partial^{\mu}\bar{\phi} 
\right.\\ 
&\qquad\qquad\qquad\qquad\left.-\frac{1}{2}\bar{m}^2\bar{\phi}^2 + \frac{1}{4}\bar{g}\bar{\phi}\tilde{f}_{\mu \nu}f^{\mu \nu}\right\}.
\end{split}\label{initialaction}
\end{equation}  Here, $f_{\mu\nu}=\partial_\mu \bar{a}_\nu-\partial_\nu \bar{a}_\mu$ stands for the electromagnetic field tensor, whereas its dual reads  
$\tilde{f}^{\mu\nu}=\frac{1}{2}\epsilon^{\mu\nu\alpha\beta}f_{\alpha\beta}$ with $\epsilon^{0123} = 1$. Hereafter, we use a metric with  
signature $\mathrm{diag}(\mathpzc{g}_{\mu \nu}) = (1,-1,-1,-1)$, and a unit system in which the speed of light, the Planck constant and the vacuum permitivity
are set to unity, $c=\hbar =\epsilon_0= 1$.  As the axion-diphoton coupling $\bar{g}=\Lambda_{\mathrm{UV}}^{-1}$  has an inverse energy 
dimension, this effective theory  belongs to the class of perturbatively  nonrenormalizable frameworks. In the following we will suppose  
that $\Lambda_{\mathrm{UV}}$ is a very large parameter in order to extend integrals over the momentum components  to an infinite-volume 
Fourier space. Whenever no problem of convergence arises, the integrals over the spacetime coordinates will also be extended to the whole 
Minkowski space.

We want to use  Eq.~(\ref{initialaction}) to derive radiative corrections up to  second-order in the axion-diphoton coupling  $\sim\bar{g}^2$. 
For this, we shall  use well-established effective field theory techniques  which have been applied extensively within the context of quantum 
gravity \cite{thooft,Stelle:1976gc,Goroff:1985sz,Goroff:1985th,Donoghue:1994dn,Donoghue:1993eb,Gies:2016con} and chiral perturbation theory \cite{s.weinberg,Gasser:1983yg,Gasser:1984gg,Ecker:1995zu} 
(see also  Refs.~\cite{Halter:1993kj,Kong:1998ic,Dicus:1997ax} for their application in nonlinear QED). In connection, we will suppose that 
$S_{\bar{g}}$ characterizes--at energies substantially lower than $\Lambda _{\mathrm{UV}}$--the leading order contribution of its UV-completion  which is Lorentz and gauge invariant. Hence, all higher-dimensional 
operators which are consistent with these fundamental symmetries  should be  included in  Eq.~(\ref{initialaction}), implying  that an infinite number of 
counterterms is necessary to cancel out the divergences linked to  one-particle irreducible Feynman diagrams \cite{Schwartz}.  However, to extract quantitative 
predictions from the radiative corrections  in the second-order approach in the axion-diphoton coupling, it is sufficient to incorporate the next-to-leading 
order term of $S_{\bar{g}}$. Combining the described method  with a dimensional analysis, we find that the renormalization of the self-energy operators 
in $\rm QED_A$  should be handled  by two local operators of dimension $6$: 
\begin{equation}
\begin{split}
&S_{\bar{g}^2}=\int d^4 x\left\{\frac{1}{2}\bar{g}^2\mathpzc{\bar{b}}_a^2 \left(\partial_\mu f^{\mu\lambda}\right)\left(\partial_\nu f^{\nu}_{\ \lambda}\right)\right.\\
&\qquad\qquad\qquad\left.+\frac{1}{2}\bar{g}^2\mathpzc{\bar{b}}_\phi^2(\partial_\mu \bar{\phi})\square(\partial^\mu\bar{\phi})+\ldots\right\}.
\end{split}\label{PLWaction}
\end{equation}Various local operators sharing both the symmetry of the theory and  the same dimensionality can be found.  However, it can be easily verified that all  
of  them reduce to those given in Eq.~(\ref{PLWaction}) through integrations  by parts. The Wilson parameters $\mathpzc{\bar{b}}_{a,\phi}$  determine  the strength of 
the contributions above. They might be determined by a matching procedure provided the UV-completion of $S_{\bar{g}}$ is known. Since we ignore the precise form of the 
latter, they will be considered  as arbitrary. We emphasize that the appearance of the square of the Wilson parameters  in $S_{\bar{g}^2}$ guarantees that--at 
least at tree level--the theory is causal [i.e. free of tachyons; see also  discussion below].

It is worth remarking that the action resulting from the combination of  Eqs.~(\ref{initialaction}) and (\ref{PLWaction}) cannot be considered as  an ordinary action  
containing higher-order derivatives. Within a quantum  effective theory approach, higher-dimensional operators--like those exhibited in Eq.~(\ref{PLWaction})--are 
suppressed by higher powers of  $\bar{g}$, so that their  consequences at low energies relative to  $\Lambda_{\rm UV}=\bar{g}^{-1}$ are tiny when compared with  the 
effects resulting directly from $S_{\bar{g}}$   \cite{Schwartz,Eichhorn:2012uv}. To all effects, they must be treated as perturbations, otherwise  a violation  of 
the unitarity  takes place  due  to  the  occurrence of  Pauli-Villars ghosts. While this problem has been noted in previous studies [see for 
instance \cite{grinstein,accioly2} and references therein], it is instructive to review it once again in the present context. To this end, we first note that the appearance 
of ghosts emerges quite straightforwardly when investigating the axion Green function that  results from combining the corresponding kinetic term and the  second line of Eq.~(\ref{PLWaction}):
\begin{equation}
\begin{split}
G(p^2)=\frac{1}{p^2-\bar{m}^2}-\frac{1}{p^2-\bar{m}_{\mathrm{s}}^2}.
\end{split}\label{Podolskyscalarpropagator}
\end{equation}At the  pole $p=\bar{m}^2$, the residue of this Green function is $\mathrm{Res}\; G(p^2)\vert_{p^2=\bar{m}^2}=1$, Although 
the exposition is made in terms of bare parameters, the idea extends straightforwardly when renormalized quantities are considered instead.
whereas at  $p^2=\bar{m}_{\rm{s}}^2=(\bar{g}\mathpzc{\bar{b}}_\phi)^{-2}$ 
it turns out to be   $\mathrm{Res}\; G(p^2)\vert_{p^2=\bar{m}_{\rm{s}}^2}=-1$. Hence, % according to the spectral representation of the corresponding propagator, 
the vacuum excitations linked to the former pole have a positive definite norm in the Hilbert space,  as should correspond to asymptotically single-particle states. Conversely, the 
square of the norm associated with the remaining  massive excitations is nonpositive and  no physical state can be associated with them  (ghost states), leading parallely to a  violation of 
unitarity \cite{grinstein}. Noteworthy, as the associated higher-dimensional operator [see Eq.~(\ref{PLWaction})] contains the square of $\bar{\mathpzc{b}}_\phi$, the ghost  
mass is real and its fictitious propagation does not involve a speed faster than the speed of light. The term written in the first line of Eq.~(\ref{PLWaction})  leads to a similar  
scenario,\footnote{The origin of the  electromagnetic theory that results  
from combining the Maxwell theory with the higher-derivative operator written in the first line of Eq.~(\ref{PLWaction}) dates  back to the work of P.~Podolsky \cite{Podolsky}. For further 
developments see Refs.~\cite{PodolskyII,Pais,Lee:1969fy,Lee:1970iw,Cutkosky:1969fq,Carlos,Accioly,Haghani}.} but with a different  ghost   mass  $\bar{m}_{\mathrm{gh}}^2=(\bar{g}\mathpzc{\bar{b}}_{a})^{-2}$ 
[for details, see Appendix~\ref{App1}]. The described situation provides evidences that symmetry arguments are  not enough to obtain  a well-behaved quantum theory  of the fields involved in  
$\mathcal{S}=S_{\bar{g}}+S_{\bar{g}^2}$. However, if the scattering matrix linked to its UV-completion is unitary,  it is natural to expect that the one associated with $\mathcal{S}$,   
covering quantum processes at lower energies $\ll\Lambda _{\mathrm{UV}}$,  is unitary too. This idea justifies the restriction given above Eq.~(\ref{Podolskyscalarpropagator}).  We remark that these  
``ghost-providing'' contributions are  redundant operators which can be dropped from the effecti\-ve Lagrangian without changing  observables \cite{Arzt:1993gz,GrosseKnetter:1993td,Georgi:1991ch}. 
Later on,  this outcome is used  to  remove them conveniently while the effective Lagrangian acquires  counter\-terms  which allow for the cancellation of the divergences associated with the 
loops that are  calculated here [see below Eq.~(\ref{genFctnal})]. 

Now,  to  carry out the renormalization program, the set of ``bare'' quantities $(\bar{m},\; \bar{g},\; \bar{m}_{\mathrm{gh}},\; \bar{m}_{\mathrm{s}},\; \bar{a}_\mu,\; \bar{\phi})$ 
should be replaced by the respective renormalized parameters $(m,\;  g,\;  m_{\mathrm{gh}},\; m_{\mathrm{s}},\; a_{\mathrm{R}}^\mu,\;\phi_{\mathrm{R}})$. In connection,  each term 
in  $\mathcal{S}=S_{\bar{g}}+S_{\bar{g}^2}+\ldots$ has to be parametrized by a renormalization constant so that the action to be considered from now on is
\begin{equation}
\begin{split}
&\mathcal{S}=\mathcal{S}_{\mathrm{R}}+\mathcal{S}_{\mathrm{ct}}+\ldots,\\
&\mathcal{S}_{\mathrm{R}}=\int d^4x \left\{ -\frac{1}{4} f_{\mathrm{R}}^2 +\frac{1}{2}(\partial\phi_\mathrm{R})^2
-\frac{1}{2}m^2 \phi_\mathrm{R}^2\right.\\
&\; \left.+\frac{1}{4}g\phi_\mathrm{R} \tilde{f}_{\mathrm{R}}f_\mathrm{R}+\frac{1}{2m_{\mathrm{gh}}^2} \left(\partial f_\mathrm{R}\right)^2-\frac{1}{2m_{\mathrm{s}}^2}\phi_\mathrm{R}\square^2\phi_\mathrm{R}\right\},\\
&\mathcal{S}_{\mathrm{ct}}=\int d^4x \left\{ -\frac{1}{4}(\mathpzc{Z}_3-1)f_{\mathrm{R}}^2 +\frac{1}{2}(\mathpzc{Z}_\phi-1) (\partial\phi_\mathrm{R})^2\right.
\end{split}\label{actionin}
\end{equation} 
\begin{equation*}
\begin{split}
&\;-\frac{1}{2}m^2 (\mathpzc{Z}_m-1) \phi_\mathrm{R}^2 +\frac{1}{2m_{\mathrm{gh}}^2}(\mathpzc{Z}_{\mathrm{gh}}-1)  \left(\partial f_\mathrm{R}\right)^2\\
&\;+\left. \frac{1}{4}(\mathpzc{Z}_{g}-1) g \phi_\mathrm{R} \tilde{f}_\mathrm{R}f_\mathrm{R}-\frac{1}{2m_{\mathrm{s}}^2}(\mathpzc{Z}_{\mathpzc{s}}-1)\phi_\mathrm{R}\square^2\phi_\mathrm{R}\right\}.
\end{split}
\end{equation*} Here,  $f_\mathrm{R}\equiv f_\mathrm{R}^{\mu\nu}=\partial^\mu a^\nu_\mathrm{R}-\partial^\nu a^\mu_\mathrm{R}$ is the renormalized electromagnetic tensor with  
$a_\mathrm{R}^\mu(x)=\mathpzc{Z}_3^{-\nicefrac{1}{2}}\bar{a}^\mu(x)$. Likewise, the renormalized 
axion field $\phi_{\mathrm{R}}(x)$ and its bare counterpart $\bar{\phi}(x)$ are connected via $\phi_{\mathrm{R}}(x)=\mathpzc{Z}_\phi^{-\nicefrac{1}{2}}\bar{\phi}(x)$, where $\mathpzc{Z}_\phi$ is the corresponding 
wavefunction renormalization constant. Any  other bare  parameter relates to its  respective renormalized  quantity  following multiplicative 
renormalizations according to 
\begin{equation}
\begin{array}{c}
\displaystyle
\bar{m}=m \sqrt{\frac{\mathpzc{Z}_{m}}{\mathpzc{Z}_{\phi}}}, \quad\bar{g}=g\frac{\mathpzc{Z}_{g}}{\mathpzc{Z}_{3}\sqrt{\mathpzc{Z}_{\phi}}},\\  \displaystyle
\bar{m}_{\mathrm{gh}}= m_{\mathrm{gh}}\sqrt{\frac{\mathpzc{Z}_{3}}{\mathpzc{Z}_{\mathrm{gh}}}},\quad \bar{m}_{\mathrm{s}}=m_{\mathrm{s}}\sqrt{\frac{\mathpzc{Z}_{\phi}}{\mathpzc{Z}_{\mathrm{s}}}}.
\end{array}\label{multiplicativerelations}
\end{equation}It is worth remarking that in Eq.~(\ref{actionin}) the shorthand notations $f_{\mathrm{R}}^2\equiv f_{\mathrm{R}\mu\nu}f_\mathrm{R}^{\mu\nu}$, 
$\tilde{f}_\mathrm{R}f_\mathrm{R}\equiv  \tilde{f}_{\mathrm{R}\mu\nu}f_\mathrm{R}^{\mu\nu}$, $(\partial \phi_\mathrm{R})^2\equiv (\partial_\mu \phi_\mathrm{R})(\partial^\mu \phi_\mathrm{R})$ 
have been used.  When inserting the expressions for the ghost masses [see below Eq.~(\ref{Podolskyscalarpropagator})] in those relations given  in the  second  line of Eq.~(\ref{multiplicativerelations}), 
we  link  the bare and renormalized  Wilsonian parameters: 
\begin{equation}
\begin{split}
\bar{\mathpzc{b}}_{a}= \mathpzc{b}_{a}\frac{1}{\mathpzc{Z}_{g}}\sqrt{\mathpzc{Z}_{3}\mathpzc{Z}_{\mathrm{gh}}\mathpzc{Z}_{\phi}},\quad \bar{\mathpzc{b}}_{\phi}= \mathpzc{b}_{\phi}\frac{\mathpzc{Z}_{3}}{\mathpzc{Z}_{g}}\sqrt{\mathpzc{Z}_{\mathrm{s}}},
\end{split}\label{wilsonianconnection}
\end{equation} where the connection between $\bar{g}$ and $g$ has been used. Noteworthy, within our second-order approximation in the coupling constant, no modification on the renormalized axion-diphoton 
coupling $g$ can be expected, i.e. $\mathpzc{Z}_{g}$ would not deviate from its classic tree-level value  $\mathpzc{Z}_{g}=1$. Hence, from now on, no distinction between the renormalized and physical coupling 
is needed. However, we emphasize that its bare counterpart $\bar{g}$ is still  subjected to a renormalization due to the wave functions renormalization constants $\mathpzc{Z}_{3}$ and $\mathpzc{Z}_{\phi}$.

At the quantum level, the dynamical information of the system described by Eq.~(\ref{actionin}) is rooted within  the Green functions. They can be obtained from  the generating functional
\begin{equation}\label{genFctnal}
Z[j,\mathpzc{j}]=\frac{\int \mathcal{D} \phi_{\mathrm{R}} \mathcal{D} a_{\mathrm{R}} \mathrm{e}^{i\mathcal{S}+i\int d^4 x  \left[-\frac{1}{2\zeta}(\partial _{\mu} a_{\mathrm{R}}^{\mu})^2+a_{\mathrm{R}}^{\mu}\mathpzc{j}_{\mu}+ \phi_{\mathrm{R}}j\right]}}{\int {\cal{D}} \phi_{\mathrm{R}} {\cal{D}} a_{\mathrm{R}} \mathrm{e}^{i\mathcal{S}}},
\end{equation}where $j(x)$ and $\mathpzc{j}^\mu(x)$ denote the external currents associated with the axion $\phi_{\mathrm{R}}(x)$ and the gauge field $a_{\mathrm{R}}^\mu(x)$.
The contribution in the exponent of  Eq.~(\ref{genFctnal}) which is proportional to the parameter $\frac{1}{\zeta}$ guarantees a covariant  quantization of $a_\mathrm{R}^\mu(x)$.  
At this point it turns out to be  convenient to bring the renormalized Lagrangian in $\mathcal{S}_{\mathrm{R}}$ [see Eq.~(\ref{actionin})] to a canonical  form in which 
the terms linked to the Pauli-Villars ghosts  are dropped.  This can be achieved by performing the following  local field redefinitions  within the path integral [see Eq.~(\ref{genFctnal})]:
\begin{equation}
\begin{split}
\phi_{\mathrm{R}}\to\phi_{\mathrm{R}}-\frac{\square}{2m_{\mathrm{s}}^2} \phi_{\mathrm{R}},\quad 
a^\mu_{\mathrm{R}}\to a^\mu_{\mathrm{R}}-\frac{\square}{2m_{\mathrm{gh}}^2} a^\mu_{\mathrm{R}},
\end{split}\label{localFR}
\end{equation}and by  keeping the accuracy to the order $g^2$. As these transformations are linear in $\phi_\mathrm{R}(x)$ and $a^\mu_{\mathrm{R}}(x)$, the associated Jacobian leads to a  decoupling  
between the  corresponding Fadeev-Popov ghosts and the fundamental fields.  The equivalence theorem \cite{Arzt:1993gz,Georgi:1991ch}  generalizes this fact by  dictating that no change is induced on the scattering 
matrix through shifts  of this  nature;  still they  modify  the initial  parameters of the theory. Indeed,  in our problem the redefinition  of $\phi_{\mathrm{R}}(x)$ leads to a kinetic term of the 
form  $\frac{1}{2}(1-\frac{m^2}{m^2_{\mathrm{s}}})(\partial\phi_\mathrm{R})^2$.  The additional factor contained in this expression  can be reabsorbed in  $\mathcal{S}_{\mathrm{ct}}$ [see  Eq.~(\ref{actionin})] 
by  redefining the wavefunction renormalization constant for the axion field  $\mathpzc{Z}_{\phi}-\frac{m^2}{m^2_{\mathrm{s}}}\to \mathpzc{Z}_{\phi}^\prime$. Similarly, we redefine 
$\mathpzc{Z}_{\mathrm{gh}}-\mathpzc{Z}_{3} \to \mathpzc{Z}_{\mathrm{gh}}^\prime$, $\mathpzc{Z}_{\mathrm{s}}-\mathpzc{Z}_{\phi} \to \mathpzc{Z}_{\mathrm{s}}^\prime$, $\mathpzc{Z}_{3}\to \mathpzc{Z}_{3}^\prime$ and 
$\mathpzc{Z}_{m}\to \mathpzc{Z}_{m}^\prime$ to reabsorb terms arising when transforming $\mathcal{S}_{\mathrm{ct}}$. Therefore, apart from  this unobservable effect, the Pauli-Villars ghosts have no result  
other than to remove the  divergences that might arise from the respective one-particle irreducible graphs.

%%%%%%%%%%%%%%%%%%%%%%%%%%%%%%%%%%%%%%%%%%%%%%%%%%%%%%%%%%%%%%%%%%%%%%%%%%%%%%%%%%%%%%%%%%%%%%%%%%%
\subsection{Renormalized  photon propagator in a modified minimal subtraction scheme\label{poMS}}
%%%%%%%%%%%%%%%%%%%%%%%%%%%%%%%%%%%%%%%%%%%%%%%%%%%%%%%%%%%%%%%%%%%%%%%%%%%%%%%%%%%%%%%%%%%%%%%%%%%

\begin{figure}
\includegraphics[width=.35\textwidth]{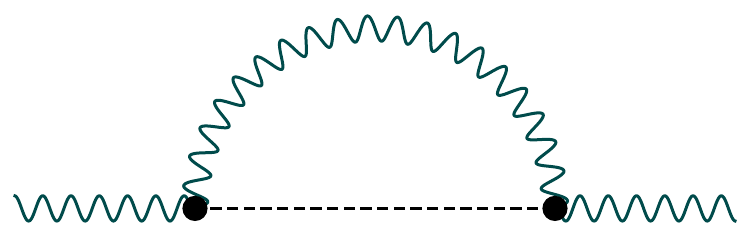}
\caption{\label{fig:mb000} Feynman diagram depicting  the axion-modified vacuum polarization tensor.  While the  dashed line represents
the free  axion propagator, the internal wavy line denotes the free photon propagator $D_{\mu\nu}^{(0)}(x,\tilde{x})$. Here, the external  
wavy lines  represent  amputated photon legs.}
\end{figure}

We pursue our investigation by determining  the  modification to  the photon propagator  $D^{(0)}_{\alpha\beta}(x,\tilde{x})$ due to quantum vacuum fluctuations of a  pseudoscalar axion field $\phi_{\mathrm{R}}(x)$. 
The use of Eq.~(\ref{genFctnal}) allows us to express the modified photon propagator  
$D_{\alpha\beta}(x,\tilde{x})$   as
\begin{equation}\label{green}
D_{\alpha \beta}(x,\tilde{x}) = \frac{1}{i^2}\left.\frac{\delta^2 Z[j,\mathpzc{j}]}{\delta \mathpzc{j}^{\alpha}(x) \delta \mathpzc{j}^{\beta}(\tilde{x})}\right|_{j,\mathpzc{j}=0}.
\end{equation}We then expand   Eq.~\eqref{genFctnal} up to the order $g^2$ and  insert the resulting expression  into the formula above. As a consequence,  the corrected  photon propagator reads
\begin{equation}\label{startingphotonpropagator}
\begin{split}
&D_{\alpha\beta }(x,\tilde{x}) = D_{\alpha \beta}^{(0)}(x,\tilde{x}) + \int d^4 y d^4 \tilde{y}\;D_{\alpha\mu}^{(0)} (x,y)\\ 
&\qquad\qquad\qquad\quad\times i\Pi^{\mu \nu}(y,\tilde{y}) D_{\nu \beta}^{(0)}(\tilde{y},\tilde{x})+\mathcal{O}(g^4).
\end{split}
\end{equation}Here  $\Pi^{\mu\nu}(y,\tilde{y})$ encompasses the  expression for the unrenormalized polarization tensor [see Fig.~\ref{fig:mb000}] as well as counterterms  that  allow for the  cancellation of the 
divergences associated with this loop.  Analytically, it reads 
\begin{equation}\label{startingVP}
\begin{split}
&\Pi^{\mu\nu}(y,\tilde{y}) =ig^2 \epsilon^{\mu \alpha \epsilon \tau} \epsilon^{\nu\beta \sigma \rho} \left[\partial_{\sigma}^{\tilde{y}} \partial_{\epsilon}^{y} \Delta_{\mathrm{F}}^{(0)}(\tilde{y},y)\right]\\
&\qquad\times  \left[ \partial_{\tau}^{y} \partial_{\rho}^{\tilde{y}} D^{(0)}_{\alpha \beta}(y,\tilde{y})\right]+\Big\{%\frac{\square}{m_{\mathrm{gh}}^2}+
\mathpzc{Z}_3^\prime-1+\frac{\square}{m_{\mathrm{gh}}^2}\\&\qquad\times(\mathpzc{Z}_{\mathrm{gh}}^\prime-1)\Big\}\left[\square\mathpzc{\mathpzc{g}}^{\mu\nu}-\partial^\mu\partial^\nu\right]\delta^4(y-\tilde{y}),
\end{split}
\end{equation}where 
$\Delta_{\mathrm{F}}^{(0)}(x,\tilde{x})%=-iG(x,\tilde{x}) 
= \int \dbar^4 p\frac{i}{p^2 -m^2 +i0}\mathrm{e}^{ip(x-\tilde{x})}$, with  $\dbar^4 p\equiv d^4p/(2\pi)^4$ refers to  the unperturbed ALP propagator, whereas  
$D^{(0)}_{\alpha\beta}(x,\tilde{x})=%-iG^{(0)}_{\alpha\beta}(x,\tilde{x}) = 
\int \dbar^4 p \frac{-i\mathpzc{g}_{\alpha\beta}}{p^2 + i0}\mathrm{e}^{ip(x-\tilde{x})}%\Big[-g_{\alpha\beta} \\ &\qquad\qquad\qquad\qquad\qquad +\left. \left( 1 - \zeta \right) \frac{p_{\alpha} p_{\beta}}{p^2} \right]
$ denotes the photon propagator  in  Feynman gauge [$\zeta=1$]. 
 It is worth remarking that, in  momentum space 
the  polarization tensor $\Pi^{\mu \nu}(p_1, p_2) = \int d^4x d^4\tilde{x} \mathrm{e}^{-ip_1x} \Pi^{\mu \nu}(x,\tilde{x})e^{ip_2\tilde{x}}$ reads
\begin{equation}
\begin{split}\label{unrenormalizedpolarizationtensor}
&\Pi^{\mu \nu}(p_1, p_2) = \deltabar_{p_1,p_2}\Big\{ ig^2\epsilon^{\mu \tau \alpha \beta} \epsilon^{\nu \sigma \gamma \rho}    g_{\sigma \tau} p_{2\beta} p_{2\rho} \mathpzc{K}_{\;\alpha\gamma}\\
&\quad -\left[%\frac{p_2^2}{m_{\mathrm{gh}}^2}+
\mathpzc{Z}_3^\prime-1-\frac{p_2^2}{m_{\mathrm{gh}}^2}(\mathpzc{Z}_{\mathrm{gh}}^\prime-1)\right](p_2^2\;\mathpzc{\mathpzc{g}}^{\mu\nu}-p_2^\mu p_2^\nu)\Big\},
\end{split}
\end{equation}where the shorthand notation $\deltabar_{p_1,p_2}\equiv(2\pi)^4\delta^4(p_1-p_2)$ has been introduced and 
\begin{equation}\label{piFourier2}
\begin{split}
\mathpzc{K}_{\;\alpha\gamma}=\int  \frac{\dbar^4q\;q_{\alpha}q_{\gamma}}{q^2 \left[ (q-p_2)^2 - m^2\right]},
\end{split}
\end{equation}which diverges quadratically as $\vert q\vert \to \infty$. The  regularization of  $\Pi_{\mu \nu}(p_1, p_2)$ is then carried out 
by using  a standard Feynman parametrization [$\frac{1}{ab} = \int_0^1 \frac{ds}{\left[ b + (a-b)s)\right]^2} $] and  by continuing  the 
loop integral to $D = 4- \epsilon$, $\epsilon \to 0^+$ dimensions via the replacement  
\begin{equation}
\int \dbar^4q\ldots \to (\mathcal{C}\mu)^{\epsilon}\int\dbar^Dq\ldots,\label{dregularization}
\end{equation} 
where $\mathcal{C}=e^{\frac{1}{2}(\gamma-1)}/(4\pi)^{\nicefrac{1}{2}}$ and   $\gamma = 0.5772...$ is  the Euler-Mascheroni constant. In this context, $\mu$ denotes 
a dimensionful parameter, i.e. the substracting point that follows when rescaling the renormalized axion-diphoton coupling $g\to g\mu^{\nicefrac{\epsilon}{2}}$  
in  $D-$dimensions so that its  mass dimension $-1$ is kept. Also, when going from four to $D-$dimensions,  the  Wilsonian parameters [see Eqs.~(\ref{PLWaction}) and (\ref{wilsonianconnection})] 
rescale  $\mathpzc{b}_i\to \mathpzc{b}_i\mu^{-\nicefrac{\epsilon}{2}}$ with $i=a,\; \phi$ while  their dimensionless feature is retained.

Now,  we integrate over $q$ and Taylor expand the resulting expression in $\epsilon$.   As a consequence, Eq.~(\ref{unrenormalizedpolarizationtensor}) becomes
\begin{equation}
\begin{split}
&\Pi^{\mu \nu}(p_1, p_2) = \deltabar_{p_1,p_2}(\mathpzc{g}^{\mu \nu} p_2^2 - p_2^{\mu} p_2^{\nu} ) \pi(p_2^2),\\
&\pi(p^2) = \frac{g^2}{16\pi^2}  
\int_0^1 ds \Delta(s) \left[ \frac{2 }{\epsilon} - \ln\left(\frac{\Delta(s)}{\mu^2}\right)\right]\\&\qquad%+\frac{p^2}{m^2_{\mathrm{gh}}}
-(\mathpzc{Z}_3^\prime-1)+\frac{p^2}{m^2_{\mathrm{gh}}}(\mathpzc{Z}_{\mathrm{gh}}^\prime-1)
\end{split}\label{unrenormalizedpolarizationtensormomentum}
\end{equation}with  $\Delta(s) = m^2s - p^2s(1-s)$. Manifestly, the term associated with the factor  $\epsilon^{-1}$ is singular as $\epsilon\to0$. In contrast to QED,  such a divergence 
cannot be  reabsorbed fully in the wavefunction renormalization constant of the electromagnetic field  $\mathpzc{Z}_3^\prime$ by enforcing that the radiative correction should not alter the residue 
of the photon propagator at $p^2=0$ \cite{Weinberg:1995mt,Schwartz}.  We solve this problem, by choosing  the counterterms in the following form  
\begin{equation}\label{counterterms}
\begin{split}
\mathpzc{Z}_3^\prime-1=\lim_{\epsilon\to0}\frac{g^2m^2}{16\pi^2\epsilon},\ \  \mathpzc{Z}_{\mathrm{gh}}^\prime-1=\lim_{\epsilon\to0}\frac{g^2 m_{\mathrm{gh}}^2}{48\pi^2\epsilon}.
\end{split}
\end{equation}Notice that the ratio  of scales $mg$ acts like a dimensionless coupling constant.  Thus,  the one-loop  renormalized polarization tensor in a modified minimal subtraction scheme  
($\overline{\mathrm{MS}}$) scheme  reads
\begin{equation}\label{polarizationMS}
\begin{split}
&\Pi^{\mu\nu}_{\overline{\mathrm{MS}}}(p_1,p_2)=\deltabar_{p_1,p_2}\left[p_2^2\mathpzc{g}^{\mu\nu}-p_2^{\mu}p_2^{\nu}\right]\pi_{\overline{\mathrm{MS}}}(p_2^2),\\
&\pi_{\overline{\mathrm{MS}}}(p^2)=-\frac{g^2}{16\pi^2}\int_0^1ds\;\Delta(s)\ln\left(\frac{\Delta(s)}{\mu^2}\right).
\end{split}
\end{equation} Noteworthy, this expression satisfies the transversality condition  $p_{1\mu}\Pi_{\overline{\mathrm{MS}}}^{\mu \nu}(p_1, p_2)=\Pi_{\overline{\mathrm{MS}}}^{\mu \nu}(p_1, p_2) p_{2\mu}=0$. 

Some comments are in order. First,  when the  QED action is extended with  those  terms belonging to   $\rm QED_A$,   the expression for $\mathpzc{Z}_3^\prime-1$ found in  Eq.~(\ref{counterterms})
will  be added  to  the corresponding  one-loop QED-expression [$\mathpzc{Z}_{3\;(\mathrm{QED})}$]\footnote{An explicit expression for  $\mathpzc{Z}_{3\;(\mathrm{QED})}$, in dimensional regularization,  
can be found  in Eq.~($19.37$) of Ref.~\cite{Schwartz}}. This operation allows us to define the  standard renormalized $\rm U(1)-$charge as usual  $e_{\mathrm{R}}=\mathpzc{\tilde{Z}}_{3\;(\mathrm{1loop})}^{\nicefrac{1}{2}}\bar{e}$, with the bare charge $\bar{e}$ 
and $\mathpzc{\tilde{Z}}_{3\;(\mathrm{1loop})}=\mathpzc{Z}_{3}^\prime+\mathpzc{Z}_{3\;(\mathrm{QED})}$. Finally, taking into account Eqs.~(\ref{counterterms}) and (\ref{polarizationMS}), 
the Fourier transform of  Eq.~(\ref{startingphotonpropagator}) is, up to an unessential longitudinal term, 
\begin{equation}\label{renormalizedpropagatorMS}
\begin{split}
&D_{\overline{\mathrm{MS}}}^{\mu\nu}(p_1,p_2)=\deltabar_{p_1,p_2} D_{\overline{\mathrm{MS}}}^{\mu\nu}(p_2), \\      &D_{\overline{\mathrm{MS}}}^{\mu\nu}(p)=\frac{-i\mathpzc{g}^{\mu\nu}}{p^2}\left[1+\pi_{\overline{\mathrm{MS}}}(p^2)\right].
\end{split}
\end{equation} This formula constitutes the starting point for further considerations.  In the next section it will be  used   to establish the axion-modified Coulomb potential.

%%%%%%%%%%%%%%%%%%%%%%%%%%%%%%%%%%%%%%%%%%%%%%%%%%%%%%%%%%%%%%%%%%%%%%%%%%%%%%%%%%%%%%%%%%%%%%%%%%%
\subsection{Axion self-energy operator, renormalized mass vs  physical mass}
%%%%%%%%%%%%%%%%%%%%%%%%%%%%%%%%%%%%%%%%%%%%%%%%%%%%%%%%%%%%%%%%%%%%%%%%%%%%%%%%%%%%%%%%%%%%%%%%%%%

\begin{figure}
\includegraphics[width=.35\textwidth]{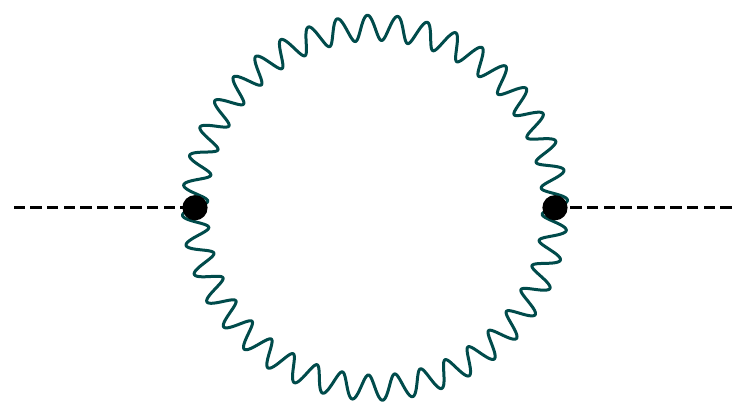}
\caption{\label{fig.001} Diagrammatic representation  of the axion  self-energy operator mediated by  quantum vacuum  fluctuations  of the electromagnetic field. In contrast to Fig.~\ref{fig:mb000}, 
the external  dashed lines represent  amputated ALPs legs, whereas the  internal wavy lines represent  photon propagators $D_{\mu\nu}^{(0)}(x,\tilde{x})$. }
\end{figure}

Our aim in this section is to determine the axion self-energy operator. Its  associated  Feynman diagram is depicted in  Fig.~\ref{fig.001}. This object  encloses 
the way in which  the quantum vacuum  fluctuations of the electromagnetic field  correct  the axion propagator. To show this analytically,  we  expand  the generating 
functional for the Green function [see Eq.~(\ref{genFctnal})] up to first order in  $g^2$.  Once this step has been  carried out, the resulting expression is twice 
differentiated functionally  with respect to the axion source $j(x)$  leading to
\begin{equation}\label{startingaxionpropagator}
\begin{split}
&\Delta_{\mathrm{F}}(x,\tilde{x}) = \frac{1}{i^2}\left.\frac{\delta^2 Z[j,\mathpzc{j}]}{\delta j(x) \delta j(\tilde{x})}\right|_{j,\mathpzc{j}=0}\\ 
&\qquad\qquad=\Delta_{\mathrm{F}}^{(0)}(x,\tilde{x}) -\int d^4 y d^4 \tilde{y}\;\Delta_{\mathrm{F}}^{(0)} (x,y)\\ 
&\qquad\qquad\qquad\qquad\times i\Sigma(y,\tilde{y}) \Delta_{\mathrm{F}}^{(0)}(\tilde{y},\tilde{x})+\mathcal{O}(g^4).\\
\end{split}
\end{equation}Here,  $\Sigma(y,\tilde{y})$ comprises the expression of the unrenormalized  axion self-energy operator as well as  some  possible counterterms. Explicitly, 
\begin{equation}\label{startingselfenergy}
\begin{split}
&\Sigma(y,\tilde{y}) =-\frac{i}{2}g^2 \epsilon^{\mu \nu \tau\sigma} \epsilon^{\alpha\beta \rho\gamma} \left[\partial_{\sigma}^{y} \partial_{\gamma}^{\tilde{y}} D_{\mu\alpha}^{(0)}(y,\tilde{y})\right]\\
&\qquad\times  \left[ \partial_{\tau}^{y} \partial_{\rho}^{\tilde{y}} D^{(0)}_{\beta \nu}(\tilde{y},y)\right]-\Big\{%\frac{\square^2}{m_{\mathrm{s}}^2}+
\square\left(\mathpzc{Z}_\phi^\prime-1\right)\\&\qquad+\frac{\square^2}{m_{\mathrm{s}}^2}(\mathpzc{Z}_{\mathrm{s}}^\prime-1)+m^2(\mathpzc{Z}_m^\prime-1)\Big\}\delta^4(y-\tilde{y}).
\end{split}
\end{equation}

Next, we Fourier transform $\Sigma(y,\tilde{y})$ and regularize its divergent integral via dimensional regularization as made in Sec.~\ref{poMS}. However, in contrast  to the case treated there,   
the  associated divergence at $\epsilon\to0$ is fully  reabsorbed  here  in  a renormalization constant 
\begin{equation}
\mathpzc{Z}_{\mathrm{s}}^\prime-1=\lim_{\epsilon\to0}\frac{g^2m_{\mathrm{s}}^2}{32\pi^2\epsilon},
\end{equation}whereas $\mathpzc{Z}_{m}^\prime$ does not deviate from its classic tree-level value  [$\mathpzc{Z}_{m}^\prime=1$]. This feature extends   beyond the one-loop approximation  because  the 
axion-photon vertex  prevents  the  proliferation  of  self-interacting terms for ALPs containing no  derivatives \cite{Eichhorn:2012uv}. Despite  this, the bare mass  $\bar{m}$ still  is subject  
to a finite renormalization  due to the axion wavefunction renormalization constant [see Eq.~(\ref{multiplicativerelations})], which  does not deviate from the classical value $\mathpzc{Z}_\phi^\prime=1$. 
Keeping in mind  all these details, we find that--in a $\overline{\mathrm{MS}}$ scheme--the renormalized axion self-energy operator  is given by
\begin{equation}
\begin{array}{c}\displaystyle
\Sigma_{\overline{\mathrm{MS}}}(p_1,p_2)=\deltabar_{p_1,p_2}\Sigma_{\overline{\mathrm{MS}}}(p_2^2),\\ \\ \displaystyle
\Sigma_{\overline{\mathrm{MS}}}(p^2)=\frac{3g^2 p^2}{32\pi^2}\int_0^1ds\; \Delta_{m\to0} (s)\ln\left(\frac{\Delta_{m\to0} (s)}{\mu^2}\right),
\end{array}\label{MSSE}
\end{equation}where $\Delta_{m\to0}(s)=-p^2s(1-s)$ [see below Eq.~(\ref{unrenormalizedpolarizationtensormomentum})].  As we could have anticipated,  this  expression is independent of the renormalized axion mass. It is,  
perhaps, worth stressing that--in a $\rm \overline{MS}-$scheme--the square of the physical mass  $m_{\mathrm{phy}}^2$ is the value of $p^2$ for which 
the real part of the two-points irreducible function:\footnote{This expression can be  established  from the identity $\int \dbar^4q\; \Gamma(p_1,q)\Delta_{\mathrm{F}}(q,p_2)=-i\deltabar_{p_1,p_2}$, where $\Delta_{\mathrm{F}}(q,p_2)$ 
stands for the Fourier transform of Eq.~(\ref{startingaxionpropagator}).} 
\begin{equation}
\begin{array}{c}\displaystyle
\Gamma(p_1,p_2)=\deltabar_{p_1,p_2}\left[p_2^2-m_{\mathrm{phy}}^2\right],\\ \\ \displaystyle m_{\mathrm{phy}}^2=m^2-\Sigma_{\overline{\mathrm{MS}}}(m_{\mathrm{phy}}^2)
\end{array}
\end{equation} vanishes. Whenever the subtracting parameter satisfies  $\mu\gg m_{\mathrm{phy}}\exp[-32\pi^2/(g^2m_{\mathrm{phy}}^2)]$,  $m_{\mathrm{phy}}\approx m$ holds. Hence, the expression for the polarization tensor  
[see Eq.~(\ref{polarizationMS})]  as a function of  $m_{\mathrm{phy}}^2$ would  not  differ  from the one given in terms of  the renormalized mass. 

At this point we find it interesting to make a comparison between Eq.~\eqref{MSSE} and the polarization tensor given in Eq.~\eqref{polarizationMS}. To this end, it is convenient to reexpress the latter 
as follows:
\begin{equation}
\begin{array}{c}\displaystyle
\Pi^{\mu\nu}_{\overline{\mathrm{MS}}}(p_1,p_2)= \deltabar_{p_1,p_2}\varkappa_{\overline{\mathrm{MS}}}(p^2) \left[\mathpzc{g}^{\mu \nu} - \frac{p_2^\mu p_2^\nu }{p_2^2}\right],\\ \\ \displaystyle
\varkappa_{\overline{\mathrm{MS}}}(p^2)=p^2\pi_{\overline{\mathrm{MS}}}(p^2).
\end{array}
\end{equation} In this formula, $\varkappa_{\overline{\mathrm{MS}}}(p^2)$  represents the only nontrivial eigenvalue of the polarization tensor \cite{Gies:2009wx,VillalbaChavez:2018ea}, which--in the limit 
under consideration--turns out to be smaller than $\Sigma_{\overline{\mathrm{MS}}}(p^2)$ by a factor $-2/3$.

Let us finally remark that, in addition to the axion-diphoton interplay, axion self-coupling \cite{cortona} as well as effective interactions with electron, proton and neutron might occur 
\cite{Ringwald:2012hr,Giannotti:2017hny,Irastorza:2018dyq,Dillon:2018ypt}. In such a case further one-loop contributions to the axion self-energy operator might arise. However, these contributions 
depend on coupling constants other than the one mediating the interaction between  an axion  and two photons.

%%%%%%%%%%%%%%%%%%%%%%%%%%%%%%%%%%%%%%%%%%%%%%%%%%%%%%%%%%%%%%%%%%%%%%%%%%%%%%%%%%%%%%%%%%%%%%%%%%%%%%%%%%%%%%%%%%%%%%%%%%%%%%%%%%%%%%%%%%%%%
\section{Axion-Coulomb potential \label{ACPsec}}
%%%%%%%%%%%%%%%%%%%%%%%%%%%%%%%%%%%%%%%%%%%%%%%%%%%%%%%%%%%%%%%%%%%%%%%%%%%%%%%%%%%%%%%%%%%%%%%%%%%%%%%%%%%%%%%%%%%%%%%%%%%%%%%%%%%%%%%%%%%%%

%%%%%%%%%%%%%%%%%%%%%%%%%%%%%%%%%%%%%%%%%%%%%%%%%%%%%%%%%%%%%%%%%%%%%%%%%%%%%%%%%%%%%%%%%%%%%%%%%%%%%%%%%%%%%%%%
\subsection{Screening of the electric charge and finite renormalization: Setting the subtracting parameter \label{ACPseca}}
%%%%%%%%%%%%%%%%%%%%%%%%%%%%%%%%%%%%%%%%%%%%%%%%%%%%%%%%%%%%%%%%%%%%%%%%%%%%%%%%%%%%%%%%%%%%%%%%%%%%%%%%%%%%%%%%

\begin{figure*}[t]
\centering
\includegraphics[width=8cm]{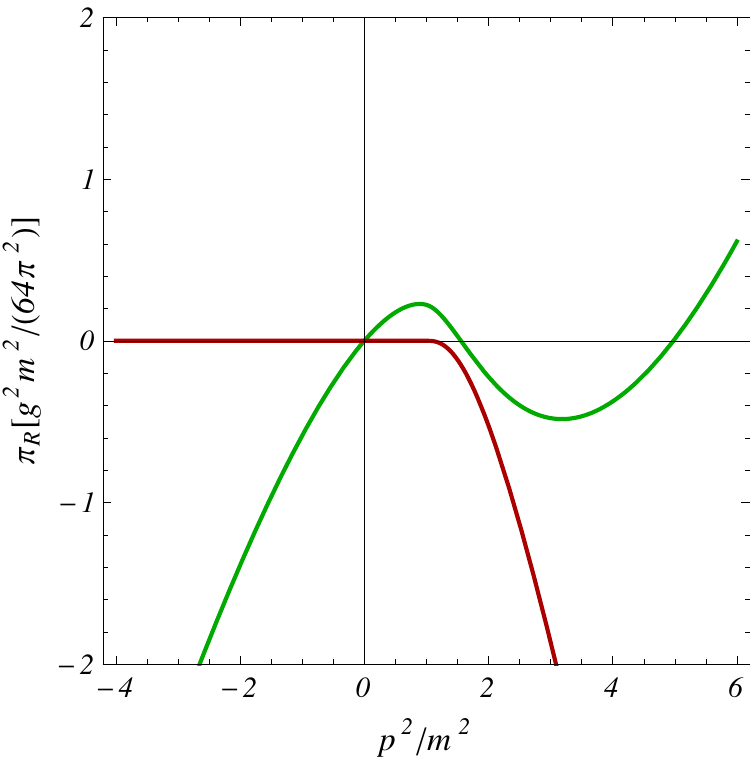}
\includegraphics[width=8cm]{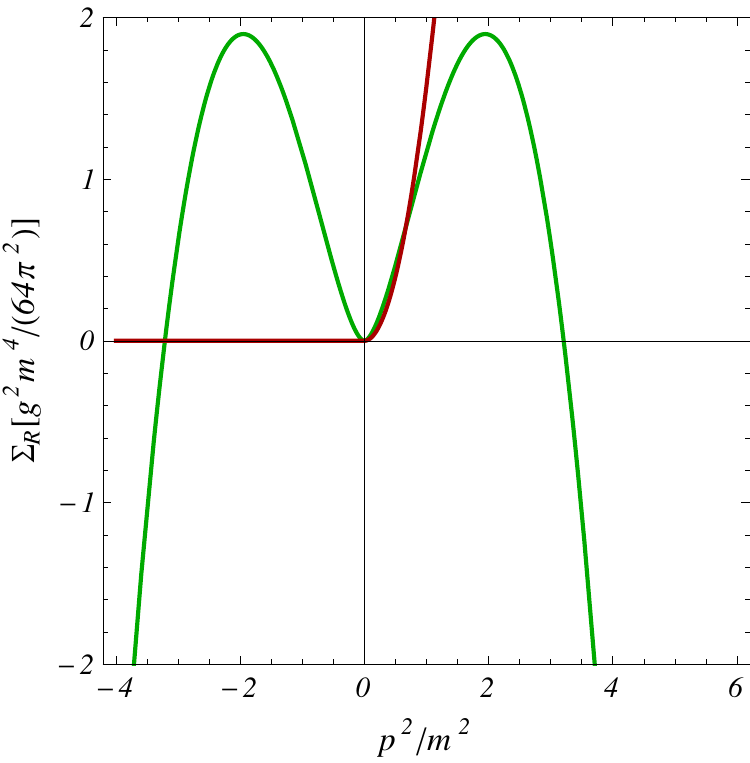}
\caption{\label{fig:mb002} In the left panel, the behavior  of the form factor of the vacuum polarization tensor  $\pi_{\mathrm{R}}(p^2)$ [see Eq.~(\ref{polarizationtensor})] is shown 
as a function of $p^2/m^2$. The right panel depicts the corresponding dependence of  the axion self-energy operator $\Sigma_{\mathrm{R}}(p^2)$ [see Eq.~(\ref{massoperator})]. The  respective real and imaginary 
parts are displayed  in green and red.}
\end{figure*}

Hypothetical  distortions of  Coulomb's law can always  be   determined  through the temporal component  of the electromagnetic four-potential [hereafter, to simpify notation, $\mathpzc{a}^\mu(\pmb{x})$ 
must be understood as $a_\mathrm{R}^\mu(\pmb{x})$]
\begin{equation}
\begin{split}
&\mathpzc{a}^\alpha(x)=-i\int D^{\alpha\beta}_{\overline{\mathrm{MS}}}(x,\tilde{x})j_\beta(\tilde{x})d^4\tilde{x},\\
&D^{\alpha\beta}_{\overline{\mathrm{MS}}}(x,\tilde{x})=\int\dbar^4p_1\dbar^4p_2e^{ip_1x}D^{\alpha\beta}_{\overline{\mathrm{MS}}}(p_1,p_2)e^{-ip_2\tilde{x}}
\end{split}\label{fourpotential}
\end{equation}
where $D^{\alpha\beta}_{\overline{\mathrm{MS}}}(p_1,p_2)$ is given in Eq.~(\ref{renormalizedpropagatorMS}). Here  $j^\beta(\tilde{x})=\mathpzc{q}\delta^\beta_{\ 0}\delta^3(\tilde{x})$ 
denotes the four-current density of a pointlike static  charge $\mathpzc{q}$ placed at the origin  $\tilde{\pmb{x}}=\pmb{0}$  of our reference frame. Particularizing the 
expression above  for  $\alpha=0$, we   end  up with 
\begin{equation}\label{temporalpotential}
\begin{split}
&\mathpzc{a}_0(\pmb{x})=\mathpzc{q}\int  \frac{\dbar^3p}{\pmb{p}^2}\left[1+\pi_{\overline{\mathrm{MS}}}(\pmb{p}^2)\right]e^{-i\pmb{p}\cdot\pmb{x}}.
\end{split}
\end{equation} At this point  it is worth  emphasizing  that, while the expression for $\pi_{\overline{\mathrm{MS}}}(\pmb{p}^2)$ [see Eq.~(\ref{polarizationMS})] 
is finite,  its dependence  on  the subtracting point $\mu$ introduces an  arbitrariness. To remove it, we  consider the expression of the 
electrostatic energy between  two electrons in momentum space $\mathpzc{U}(\pmb{p})$. It can be established easily by taking  the integrand above,   
with   $\mathpzc{q}\to e_{\rm R}$. After multiplying the resulting expression  by $e_{\rm R}$, we find
\begin{equation}
\mathpzc{U}(\pmb{p})=-e_{\mathrm{R}}\mathpzc{a}_0(\pmb{p})=-\frac{e_{\mathrm{scr}}^2(\pmb{p})}{\pmb{p}^2},\label{fourpotentialforiertransformed}
\end{equation}where the  screened charge $e_{\mathrm{scr}}^2(\pmb{p})=e_{\mathrm{R}}^2[1+\pi_{\overline{\mathrm{MS}}}(\pmb{p}^2)]$ has been defined.

As we still  have  freedom of performing finite renormalizations, we can  demand that  $\pi_{\overline{\mathrm{MS}}}(\pmb{p}^2)$  
vanishes as $\vert\pmb{p}\vert\to0$. Since  the corresponding length scale $\vert\pmb{x}\vert\to\infty$, $e_{\mathrm{scr}}^2(\vert\pmb{x}\vert\to\infty)$  can 
be identified with the  electrostatic charge  that  is  measured  in experiments at low energies. This natural  renormalization condition 
[$\pi_{\overline{\mathrm{MS}}}(\pmb{0})=0$] holds for  the subtracting parameter  $\mu=m\exp[-1/4]$. The renormalized polarization tensor then reads
\begin{equation}\label{polarizationtensor}
\begin{split}
&\Pi^{\alpha\beta}_{\mathrm{R}}(p_1,p_2)=\deltabar_{p_1,p_2}\left[p_2^2\mathpzc{g}^{\alpha\beta}-p_2^{\alpha}p_2^{\beta}\right]\pi_{\mathrm{R}}(p_2^2),\\
&\pi_{\mathrm{R}}(p^2)=-\frac{g^2 m^2}{64\pi^2}\left[1-\frac{1}{3}\frac{p^2}{m^2}\right]\\&\qquad\qquad\qquad-\frac{g^2}{16\pi^2}\int_0^1ds\;\Delta(s)\ln\left(\frac{\Delta(s)}{m^2}\right),
\end{split}
\end{equation}whereas the axion self-energy operator [see Eq.~(\ref{MSSE})] reduces to 
\begin{equation}\label{massoperator}
\begin{array}{c}
\displaystyle\Sigma_{\mathrm{R}}(p_1,p_2)=\deltabar_{p_1,p_2}\Sigma_{\mathrm{R}}(p_2^2),\\ \\
\displaystyle\Sigma_{\mathrm{R}}(p^2)=-\frac{g^2 p^4}{64\pi^2}\left[\ln\left(-\frac{p^2}{m^2}\right)-\frac{7}{6}\right].
\end{array}
\end{equation}

The results  obtained so far  are  summarized in  Fig.~\ref{fig:mb002}, which displays the behavior of  $\pi_{\mathrm{R}}(p^2)$ [left panel]  and $\Sigma_{\mathrm{R}}(p^2)$ [right panel] as a function  of $p^2/m^2$ .  
In both panels the respective  real and imaginary parts  are shown in green and red,   manifesting   by themselves   the non-Hermitian feature of the polarization tensor and the axion self-energy operator. To support this 
numerical evaluation from an analytic viewpoint, we first determine an exact expression for the imaginary parts. To this end, we restore the 
$i0-$prescription  [$m^2\to m^2-i0$] in Eq.~(\ref{polarizationtensor}) and apply the formula $\ln(-\mathcal{A}-i0)=\ln(\vert\mathcal{A}\vert)-i\pi$ with $\mathcal{A}>0$. Explicitly,
\begin{equation}\label{imaginarypart}
\begin{array}{c}\displaystyle
\mathrm{Im}[\pi_{\mathrm{R}}(p^2)]=-\frac{g^2p^2}{96\pi}\left[1-\frac{m^2}{p^2}\right]^3\Theta(p^2-m^2),\\ \\\displaystyle
\mathrm{Im}[\Sigma_{\mathrm{R}}(p^2)]=\frac{g^2(p^2)^2}{64\pi}\Theta(p^2-m^2),\\
\end{array}
\end{equation}where $\Theta(x)$ denotes the unit step function. %It is worth remarking that this expression  depend  neither  on the substracting point $\mu$ nor on the renormalized ghost mass $m_{\mathrm{g}}$. 
We note that for an on-shell photon  [$p^2=0$], the imaginary part of  $\Pi_{\mathrm{R}}^{\mu\nu}$ vanishes, which implies--according to the optical theorem--that 
the emission of an ALP from a photon accompanied by the radiation of another photon is forbidden. This fact agrees with the outcome resulting from an analysis 
of the corresponding energy-momentum balance.  

The expression for the imaginary part of the one-loop self-energy operator $\mathrm{Im}[\Sigma_{\mathrm{R}}(p^2)]$ [second line in Eq.~(\ref{imaginarypart})]
coincides with the result found  previously in Ref.~\cite{Gabrielli:2006im} through a direct application of the cutting rules. Its on-shell evaluation 
[$p^2=m^2$] should allow us to determine the total  rate of the decay process $\phi\to\gamma\gamma$ via the  relation 
$\mathrm{Im}[\Sigma_{\mathrm{R}}(m^2)]=m\Gamma_{\phi\to\gamma\gamma}$,  provided the optical theorem is valid. With accuracy to first order in $g^2$, this 
formula is indeed verified  because an expression for $\Gamma_{\phi\to\gamma\gamma}$--relying on the corresponding $S-$matrix amplitude--can 
be inferred directly from the corresponding neutral pion decay rate [see for instance Eq.~(19.119) in Ref.~\cite{peskin}]. This fact evidences that the unitarity  
is preserved,  at least within the second order approximation in the  axion-diphoton coupling $g$. 

Further asymptotic  expressions of   Eq.~(\ref{polarizationtensor}) are  elucidated. For $m^2\gg p^2$, we find that $\pi_{\mathrm{R}}(p^2)$  approaches to 
\begin{equation}
\pi_{\mathrm{R}}(p^2\ll m^2)\approx\frac{g^2p^2}{144\pi^2}\left[1-\frac{3}{8}\frac{p^2}{m^2}\right].
\label{smallbehaviorMSscheme}
\end{equation}Conversely, for  $\vert p^2\vert\gg m^2$,  its asymptotic behavior turns out to be dominated by the following  function
\begin{equation}
\begin{split}
&\pi_{\mathrm{R}}\vert_{\vert p^2\vert \gg m^2}\approx \frac{g^2p^2}{96\pi^2}\left[\ln\left(\frac{\vert p^2\vert}{m^2}\right)-i\pi\Theta(p^2)-\frac{7}{6}\right].
\end{split}\label{largemomentaMS}
\end{equation}Notably, when  $p$ is timelike [$p^2>0$], the expression above gets an imaginary contribution $\mathrm{Im}\left[\pi_{\mathrm{R}}(p^2\gg m^2)\right]\approx - g^2p^2/(96\pi)$, 
which coincides with the leading order term  of  Eq.~(\ref{imaginarypart}) when  the condition $p^2\gg m^2$ is considered. 

%%%%%%%%%%%%%%%%%%%%%%%%%%%%%%%%%%%%%%%%%%%%%%%%%%%%%%%%%%%%%%%%%%%%%%%%%%%%%%%%%%%%%%%%%%%%%%%%%%%%%%%%%%%%%%%%
\subsection{Electrostatic potential and  modified Lamb-shift \label{ACPsecb}}
%%%%%%%%%%%%%%%%%%%%%%%%%%%%%%%%%%%%%%%%%%%%%%%%%%%%%%%%%%%%%%%%%%%%%%%%%%%%%%%%%%%%%%%%%%%%%%%%%%%%%%%%%%%%%%%%

The first contribution in  Eq.~(\ref{temporalpotential}) can be integrated straightforwardly, leading to the unperturbed Coulomb potential  $\mathpzc{a}_C(\pmb{x})=\mathpzc{q}/(4\pi\vert\pmb{x}\vert)$. 
The second one, on the other hand, will  be computed by using $\pi_{\mathrm{R}}(\pmb{p}^2)$ rather than $\pi_{\overline{\mathrm{MS}}}(\pmb{p}^2)$. With all these details in mind we write 
\begin{equation}
\begin{split}
&\mathpzc{a}_0(\pmb{x})=\mathpzc{a}_C(\pmb{x})+\delta\mathpzc{a}(\pmb{x}),\\
&\delta\mathpzc{a}(\pmb{x}) =\mathpzc{q}\int  \frac{\dbar^3p}{\pmb{p}^2}\pi_{\mathrm{R}}(\pmb{p}^2)e^{-i\pmb{p}\cdot\pmb{x}}.
\end{split}\label{correctionCP}
\end{equation}For  evaluating $\delta\mathpzc{a}(\pmb{x})$ explicitly, it is   convenient to  integrate by parts in Eq.~(\ref{polarizationtensor}) and use an equivalent 
representation of $\pi_{\mathrm{R}}(\pmb{p}^2)$ instead:
\begin{equation}\label{polarizationtensorv2}
\begin{split}
&\pi_{\mathrm{R}}(p^2)=\frac{g^2p^2}{144\pi^2}\left[1-\frac{3}{2}p^2\int_0^1\frac{ds\;s^3}{m^2-p^2(1-s)}\right].
\end{split}
\end{equation}Observe that, for applying this formula in Eq.~(\ref{correctionCP}),  $p_0$ must be  set to zero. Taking this  into account, the integral over the momentum can be carried out  with  relative ease. After  developing the change of variable $u=1/(1-s)^{\nicefrac{1}{2}}$, 
the axion-modified potential turns out to be 
\begin{equation}
\begin{split}
&\mathpzc{a}_0(\pmb{x})=\frac{\mathpzc{q}}{4\pi\vert\pmb{x}\vert}\Big\{1+\frac{g^2m^2}{48\pi^2}\int_{1}^\infty \frac{du}{u^5}\\ &\qquad\qquad\qquad\qquad\qquad\times\left[u^2-1\right]^3e^{-m\vert\pmb{x}\vert u}\Big\}.
\end{split}
\label{generalpotenatiallwithoutb}
\end{equation}We remark that spurious contributions containing Dirac deltas $\delta^3(x)$ have been ignored since the theory is predictive only for distances  $\vert\pmb{x}\vert\gg g$. Although the integral involved 
in this formula can be calculated analytically, we will keep it as it stands. Mainly, because it will allow us to establish compact expressions for the energy shifts that atomic transitions undergo.

Asymptotic formulas for the modified potential can be extracted from Eq.~(\ref{generalpotenatiallwithoutb})  without much efforts. For instance, at distances larger than the Compton wavelength $\lambda\sim m^{-1}$ 
of the axion, i.e. for $\vert\pmb{x}\vert\gg\lambda$, the region $u\sim 1$ dominates in the integral involved in Eq.~(\ref{generalpotenatiallwithoutb}), and the axion-modified Coulomb potential 
approaches to 
\begin{equation}
\mathpzc{a}_0(\vert\pmb{x}\vert\gg\lambda)\approx\frac{\mathpzc{q}}{4\pi\vert\pmb{x}\vert}\left[1+\frac{g^2m^2}{\pi^2}\frac{e^{-m\vert\pmb{x}\vert}}{(m\vert\pmb{x}\vert)^4}\right].   
\end{equation}However, at short distances [$\vert\pmb{x}\vert\ll\lambda$], the main contribution to the integral in Eq.~(\ref{generalpotenatiallwithoutb}) results from the 
region $1\leqslant u\leqslant(m\vert\pmb{x}\vert)^{-1}$, and the integrand can be approached by its most slowly decreasing function in $u$, which is  $\sim u e^{-m\vert\pmb{x}\vert u}$. Consequently,
\begin{equation}
\begin{split}
&\mathpzc{a}_0(\vert\pmb{x}\vert\ll\lambda)\approx\frac{\mathpzc{q}}{4\pi\vert\pmb{x}\vert}\Big\{1+\frac{g^2m^2}{48\pi^2}\frac{1}{(m\vert\pmb{x}\vert)^2}\Big\}.
\end{split}
\end{equation}This expression is independent of the axion mass. Observe that the distance $\vert\pmb{x}\vert$ must satisfy the condition  $\vert\pmb{x}\vert\gg g/(4\sqrt{3}\pi)$, otherwise our 
perturbative approach breaks down. Incidentally, the corresponding energy scale  $\mu_\mathrm{p}\sim \vert\pmb{x}\vert^{-1}\sim 4\sqrt{3}\pi/g$ coincides--up to a  numerical factor of the order of 
one--with the Landau pole linked to $\rm QED_A$:  $\mu_{\mathrm{L}}\approx 4\sqrt{6}\pi/g$  [for details see  Ref.~\cite{Eichhorn:2012uv}].

\begin{figure}
\includegraphics[width=.40\textwidth]{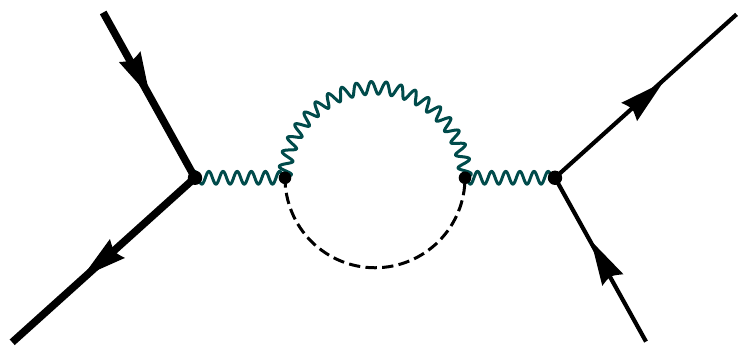}
\caption{\label{fig.003} Pictorial correction to the  Coulomb potential due to quantum vacuum fluctuations of axion  and electromagnetic fields.  Leaving aside the  electron (proton) legs 
[external lines with arrows in the right (left)], the remaining pieces of this diagram are  described in  Fig.~\ref{fig:mb000}. }
\end{figure} 

The distortion of the  Coulomb potential due to ALPs [see Eq.~(\ref{generalpotenatiallwithoutb})] allows us to infer the induced modifications in the spectrum of a nonrelativistic 
hydrogenlike atom. Since so far no large deviations from the standard QED predictions have been observed, we will assume that these energy shifts are very small and, consequently,  
apply standard time-independent perturbation theory. When considering a first order approximation, the energy shift $\delta \varepsilon$  follows  by  averaging the correction to 
the electrostatic energy  $\delta\mathpzc{U}(\pmb{x})=-e_{\mathrm{R}}\delta\mathpzc{a}(\pmb{x})$ [see Eq.~(\ref{correctionCP}) and the Feynman diagram depicted in Fig.~\ref{fig.003}] over the $0$th 
order  wavefunctions $\vert\psi_{n,\ell,j} \rangle$. Explicitly,
\begin{equation}
\label{averageenergy}
\delta \varepsilon_{n,\ell,j}^{(1)}=\langle \psi_{n,\ell,j} \mid \delta \mathpzc{U}(\pmb{x})\mid \psi_{n,\ell,j} \rangle.
\end{equation}We wish to exploit this formula to predict a plausible  axion Lamb-shift for the $2s_{1/2}-2p_{1/2}$ transition in atomic hydrogen. 
For this case,  Eq.~\eqref{averageenergy} leads to
\begin{equation}
\begin{split}
&\delta \varepsilon =\delta\varepsilon_{2s_{\nicefrac{1}{2}}}^{(1)}-\delta\varepsilon_{2p_{\nicefrac{1}{2}}}^{(1)},\\
&\qquad=\int_0^{\infty} dr\, r^2\delta \mathpzc{U}(r)   \left[R_{2s}^2(r) - R_{2p}^2(r)\right],
\end{split}
\label{lambshift}
\end{equation}where $r\equiv\vert\pmb{x}\vert$ and  $R_{n\ell}$ stands for a radial hydrogen wave function. In particular,
\begin{equation}\label{radial2s2p}
\begin{array}{c}\displaystyle
R_{2s}(r) = \frac{1}{\sqrt{2}} \frac{1}{a_{\mathrm{B}}^{3/2}}\left[1-\frac{r}{2a_{\mathrm{B}}}\right]e^{-\frac{r}{2a_{\mathrm{B}}}},\\ 
\displaystyle R_{2p}(r) = \frac{1}{2\sqrt{6}}\frac{r}{a_{\mathrm{B}}^{\nicefrac{5}{2}}} e^{-\frac{r}{2a_{\mathrm{B}}}}.
\end{array}
\end{equation}Here $a_{\mathrm{B}}=(\alpha m_e)^{-1}$ is the Bohr radius with $\alpha=1/137$ the fine structure constant and $m_e=0.511\ \rm MeV$ the electron mass.

\begin{figure}[t]
\centering
\includegraphics[width=8cm]{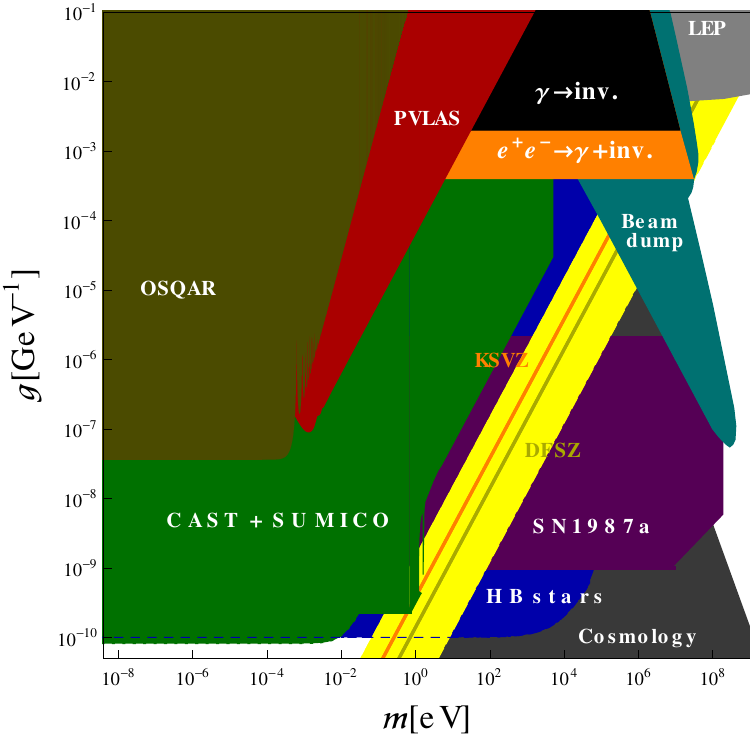}
\caption{\label{fig:mb005}Summary of exclusion areas for a pseudoscalar ALP coupled to two photons.  Compilation adapted from Refs.~\cite{Jaeckel:2010ni,Ringwald:2012hr,Hewett:2012ns,Essig:2013lka,Alekhin:2015byh}. 
The picture includes [inclined yellow band] the predictions of the axion models $\vert E/N-1.95\vert=0.07-7$ (the notation of this formula is in accordance with Ref. \cite{CAST}).   Colored in orange and black appear the regions ruled out by particle decay experiments.  
While the  portion discarded  by investigating the energy loss in the  horizontal branch  (HB) stars are shown in blue, the excluded area resulting from the solar monitoring of a plausible ALP flux (CAST+SUMICO) has been 
added in green.  In purple the portion discarded by measuring the  duration of the neutrino signal of  the supernova $\rm SN1987A$ is  depicted, whereas the dark gray area results from cosmological 
studies. The excluded area in dark turquoise has been established from  beam dump experiments. Besides, the  light gray zone  has been excluded from  electron-positron collider (LEP) investigations.
Finally, the colored sectors in olive and red show the exclusion regions corresponding to OSCAR and PVLAS collaborations, respectively. }
\end{figure}

\begin{figure*}[t]
\centering
\includegraphics[width=8cm]{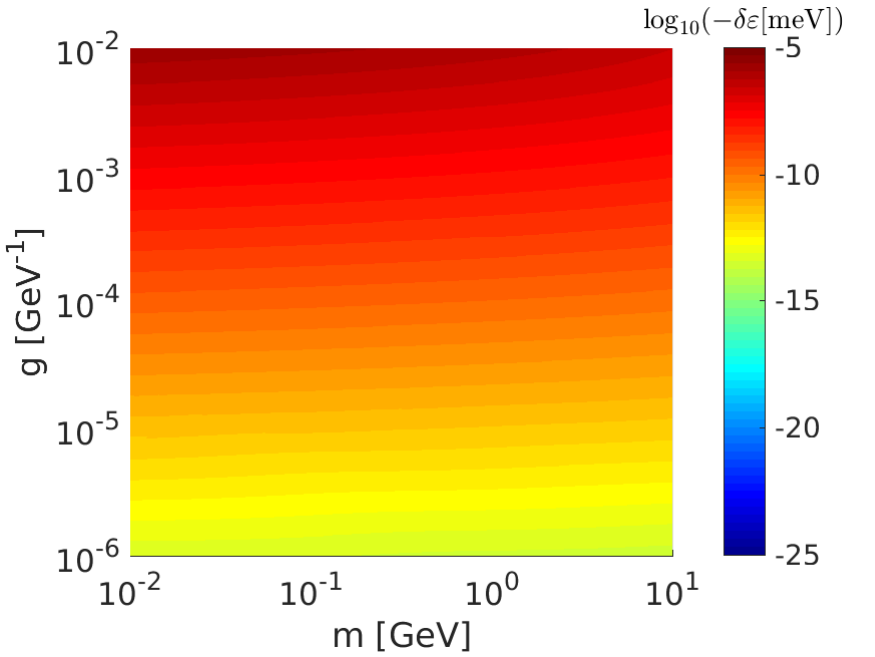}
\includegraphics[width=8cm]{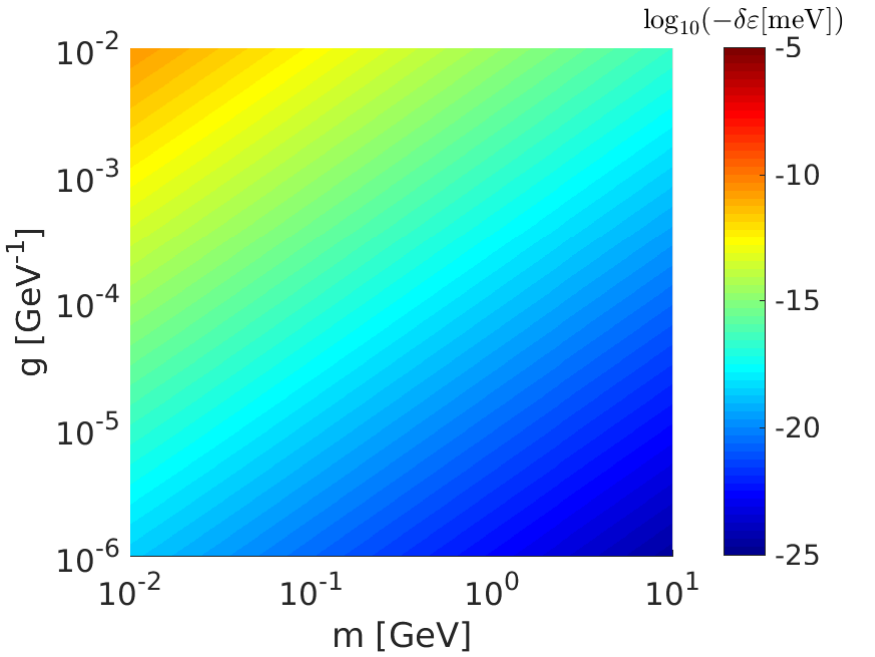}
\caption{\label{fig:mb003} Energy-shift induced by hypothetical quantum vacuum fluctuations of axionlike fields on the  $2s_{1/2}-2p_{1/2}$(left panel) and the $3p_{1/2}-3d_{1/2}$(right panel) atomic transitions in 
$\rm H_\mu$. The ALPs parameters used in these numerical evaluations are still undiscarded. While the results exhibited in the left panel rely on  Eq.~(\ref{axionlambshiftfinal}), the outcomes depicted in the 
right panel follow from Eq.~(\ref{axionlambshiftfinalfd}). Observe that the energy shift is plotted in the form of $\log_{10}(-\delta\varepsilon[\rm meV])$.}
\end{figure*}

Observe that the integration in Eq.~(\ref{lambshift}) covers the region $[0,\infty)$. However, since $\rm QED_A$ does not provide a precise information about the form of  
the axion-Coulomb potential for distances smaller than  $g$,  the integral over $r$ must be splitted  $ \int_0^{\infty} dr\ldots= \int_0^{g} dr\ldots+ \int_{ g}^{\infty} dr\ldots$. 
In the following we will assume that the contribution  from the outer region $[g,\infty)$ dominates over the  inner region  $[0,g]$, which we  ignore.\footnote{Strictly speaking, in accordance with  the 
treatment applied in Sec.~\ref{sec:QKA} [read also below Eq.~(\ref{generalpotenatiallwithoutb})], the splitting of the integral should be carried out at a certain point $d$ fulfilling 
the condition $d\gg g$.  However, in order to avoid  uncertainties stemming from this additional parameter, we set  $d=g$. Our corresponding results should be considered as order-of-magnitude estimates, accordingly.} 
As we will see very shortly, the yet undiscarded values for $g$ turn out to be much smaller than any characteristic atomic scale.  With 
all these details in mind we integrate over $r$ and arrive at
\begin{equation}
\begin{split}
&\delta \varepsilon\approx -\frac{\alpha g^2}{96\pi^2}m^4a_{\mathrm{B}}\int_1^\infty \frac{du}{u^3}\frac{[u^2-1]^3}{[1+u a_{\mathrm{B}} m]^4}e^{-gm u}.
\end{split}
\label{axionlambshiftfinal}
\end{equation} Let us study  the asymptotes of this expression. We first consider the case in  which $a_{\mathrm{B}}\gg \lambda$.  Under this condition, the term  of the 
integrand $[1+u a_{\mathrm{B}} m]^4$ is dominated by  $u^4 a_{\mathrm{B}}^4 m^4$. The integral resulting from this approximation can be computed exactly. After a Taylor expansion
in  $mg\ll1$, we find the compact expression 
\begin{equation}
\begin{split}
&\delta \varepsilon\approx \frac{\alpha g^2}{96\pi^2a_{\mathrm{B}}^3}\left[\ln\left(gm\right)+\gamma+\frac{11}{12}\right]. \label{axionlambshift1}
\end{split}
\end{equation} As in Eq.~(\ref{dregularization}),  $\gamma=0.5772\ldots$ refers to  the Euler-Mascheroni  constant. In the opposite case  $a_{\mathrm{B}}\ll\lambda$, the  integrand in 
Eq.~(\ref{axionlambshiftfinal})   turns out to be a function that decreases monotonically with the growing of  $u$. It is then justified to approach it through  its  
most slowly decreasing part which is  $\sim u^3 e^{-mgu}/(1+u a_{\mathrm{B}} m)^4$. As a result, the corresponding integral can be computed 
analytically by using (3.353.1) in Ref.~\cite{Gradshteyn}. In the limit of  $mg\ll1$, it allows us to approach 
\begin{equation}
\begin{split}
&\delta \varepsilon\approx \frac{\alpha g^2}{96\pi^2a_{\mathrm{B}}^3}\left[\ln\left(\frac{g}{a_{\mathrm{B}}}\right)+\gamma+\frac{11}{6}\right]. \label{axionlambshift2}
\end{split}
\end{equation}   Notice that, the equation above is a good approximation whenever the condition $a_{\mathrm{B}} \gg g$ holds. Moreover, although Eqs.~(\ref{axionlambshiftfinal})-(\ref{axionlambshift2})  
apply for ordinary atomic hydrogen, they can be adapted conveniently for studying the same transition in other  hydrogenlike atoms. When hydrogenlike ions with atomic number $Z>1$ are considered, for instance, the correction 
to the Lamb-shift will be given by Eqs.~(\ref{axionlambshiftfinal})-(\ref{axionlambshift2}), scaled by the factor $Z$ and  $a_{\mathrm{B}}\to a_{\mathrm{B}}/Z$. If a muonic hydrogen atom is investigated instead, a replacement 
of the electron mass $m_e$ by the reduced mass of the system  $m_r\approx186\; m_e$  would be required.

%%%%%%%%%%%%%%%%%%%%%%%%%%%%%%%%%%%%%%%%%%%%%%%%%%%%%%%%%%%%%%%%%%%%%%%%%%%%%%%%%%%%%%%%%%%%%%%%%%%%%%%%
\subsection{Precision spectroscopy in $\bf H_{\pmb \mu}$ and the proton radius anomaly \label{ACPsecc}}
%%%%%%%%%%%%%%%%%%%%%%%%%%%%%%%%%%%%%%%%%%%%%%%%%%%%%%%%%%%%%%%%%%%%%%%%%%%%%%%%%%%%%%%%%%%%%%%%%%%%%%%%

Before continuing with the physics of virtual  ALPs, we will  estimate the contribution to the Lamb-shift by a meson whose interaction with the electromagnetic field resembles the one exhibited by 
ALPs [see Eq.~(\ref{initialaction})], i.e.  the neutral pion $\pi^0$.  We should however emphasize that its  effect should be understood as a consequence of the quantum vacuum fluctuations of its 
constituent quark fields.  When thinking of the axion as $\pi^0$,  $m\to m_\pi=135 \rm \; MeV$ and  the coupling constant  $g\to\alpha/(\pi f_\pi)$ turns out to be determined by  $\alpha$ and the 
pion decay constant  $f_\pi\approx92\rm\; MeV$ \cite{Schwartz}.  Observe that the corresponding value of $g\approx 4.97\times  10^{-3}\ \rm fm$ is two orders of magnitude smaller than the proton radius 
$r_{\mathrm{proton}}\approx0.876\ \rm fm$ \cite{Pohl}. The corresponding  correction to the $2s_{1/2}-2p_{1/2}$ Lamb-shift in hydrogen atoms  is $\delta\varepsilon=-1.07\times 10^{-12}\rm\; meV$. This value turns out to be five
orders of magnitude smaller than  the experimental uncertainty $\vert\delta\varepsilon_{2\sigma}\vert=2\times 10^{-7}\ \rm meV$, established  at $2\sigma$ confidence level \cite{Glueck,Roy}. 
If the previous evaluation is carried out by considering a muonic hydrogen instead [$m_e\to m_r$], we find that the correction to the energy due to the neutral pion field 
is $\delta\varepsilon=-6.81\times 10^{-6}\rm\; meV$. Since this is five orders of magnitude smaller than the existing discrepancy between the experimental measurement and the theoretical prediction  
$\delta \varepsilon=0.31\ \rm meV$ \cite{Pohl,barger,Jentschura:2014yla},  virtual neutral pions are excluded as possible explanation for the proton radius puzzle.

Now, we wish to investigate whether the Lamb-shift induced by quantum vacuum fluctuations of axionlike fields  might cure this  anomaly. To this end, we will evaluate  Eq.~\eqref{axionlambshiftfinal} 
considering a mass region  in  which reliable results can be extracted. Within a pure spectroscopy context, this occurs for ALP wavelengths smaller or of  the order of the  Bohr radius of $\rm H_\mu$, 
i.e. $\lambda\lesssim a_\mu$ with  $a_\mu\approx285\ \rm fm$, otherwise interactions of other nature must be included. Correspondingly, we can formally explore ALP masses fulfilling the condition 
$1\ \mathrm{MeV}\lesssim m$.  However,  in the range   $1\ \mathrm{MeV}\lesssim m\lesssim 100\  \mathrm{MeV}$  the axion-diphoton coupling $g$ has been constrained severely from various 
results, including those dealing with electron beam fixed-target setups [see compilation of bounds in Refs.~\cite{Dobrich:2015jyk,Jaeckel:2015jla}]. Conversely, the sensitivities in experiments  
where ALPs masses  $100\ \mathrm{MeV}\lesssim m\lesssim 10\  \mathrm{GeV}$ are probed turn out to be much weaker [white sector in the right hand side of Fig.~\ref{fig:mb005}]. A recent 
investigation based on electron-positron colliders has constrained $g$ to lie below $g<10^{-2}\ \rm GeV^{-1}$ \cite{Jaeckel:2015jla}.

A numerical assessment of the axion-modified Lamb-shift has been carried out by considering this  yet undiscarded region. The outcome of this evaluation  is summarized in the left panel of 
Fig.~\ref{fig:mb003}.  Observe that the energy shift has been  plotted in the form of $\log_{10}(-\delta\varepsilon[\rm meV])$. The highest value achieved for $\delta\varepsilon\sim-10^{-6}\ \rm meV$ 
corresponds to  $g\sim 10^{-2}\ \rm GeV^{-1}$ and $m\sim10^{-2}\ \rm GeV$. Toward higher axion masses $m\sim 10\ \rm GeV$ and lower axion-diphoton couplings $g\sim 10^{-6}\ \rm GeV^{-1}$ the correction 
to the Lamb-shift tends to decrease significantly [$\delta\varepsilon\sim - 10^{-14}\ \rm meV$]. Both estimates coincide  with the  values resulting from 
Eq.~(\ref{axionlambshift1}). The smallness of these outcomes as compared with the aforementioned discrepancy rules out the corresponding virtual ALPs as candidates to explain the $\rm H_\mu$  anomaly.  
As we have anticipated above Eq.~(\ref{axionlambshiftfinal}), the chosen values for  $10^{-7}\ \mathrm{fm}\lesssim g \lesssim 10^{-4}\ \rm fm$ are much smaller than $r_{\mathrm{proton}}\approx0.876\ \rm fm$.

\begin{figure*}[t]
\centering
\includegraphics[width=8cm]{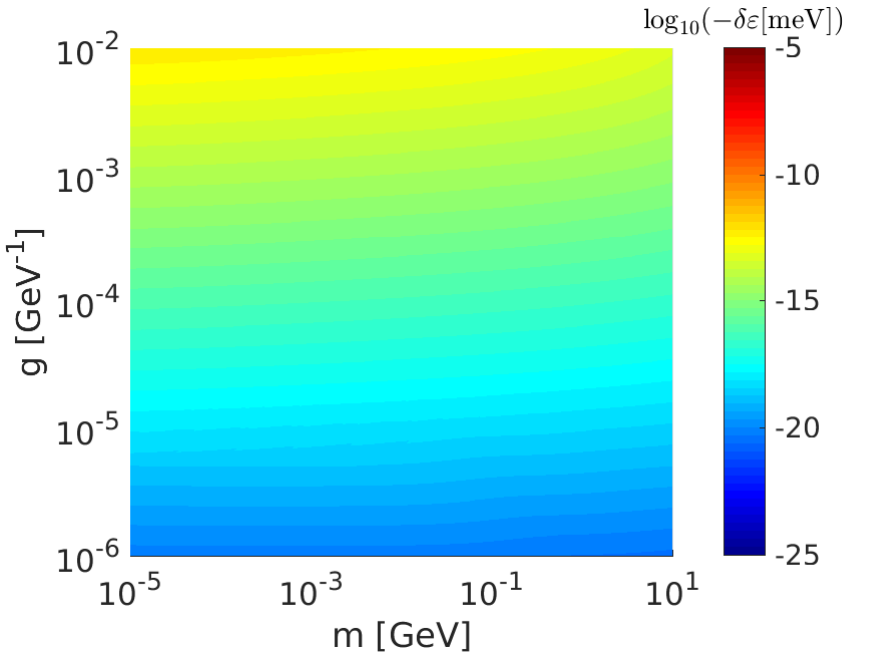}
\includegraphics[width=8cm]{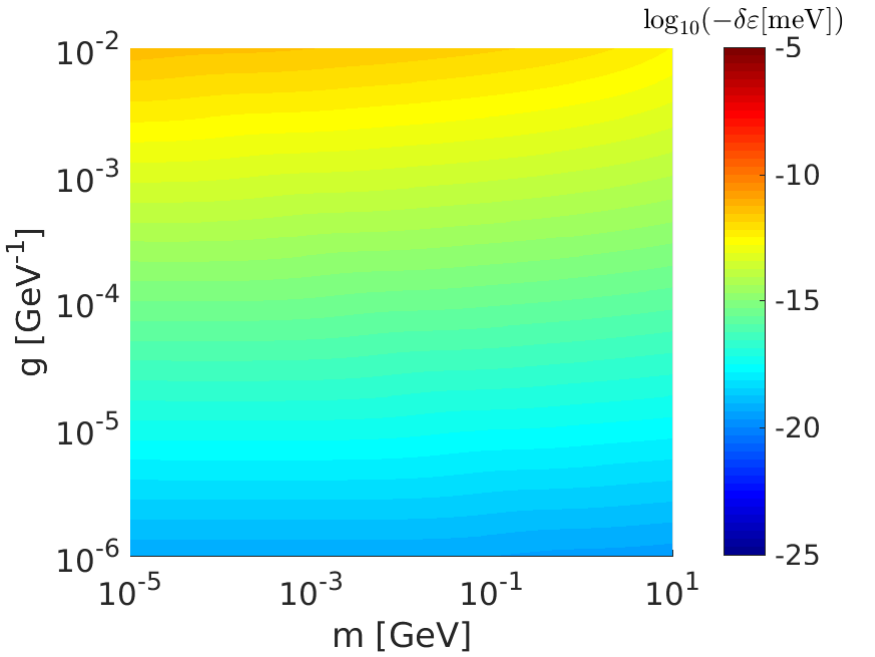}
\caption{\label{fig:mb004} Projected sensitivities for the $2s_{1/2}-2p_{1/2}$ [see Eq.~\eqref{axionlambshiftfinal}] and $1s_{1/2}-2s_{1/2}$ [see Eq.~\eqref{axionlambshiftathydrogen}] atomic transitions 
in hydrogen are depicted in the left and the right panel, respectively. In  contrast to  Fig.~\ref{fig:mb003}, the region for the axion mass $m$ is wider, covering  from $10^{-5}\ \rm GeV$ to $10\  \rm GeV$.
As previously,  $10^{-6}\ \mathrm{GeV}^{-1}<g<10^{2}\ \rm GeV^{-1}$  and the energy-shift is given in the  form of $\log_{10}(-\delta\varepsilon[\rm meV])$.}
\end{figure*}

Clearly, the previous statements cannot be considered conclusive  as our estimation undergoes theoretical uncertainties arising from both the internal limitation of $\rm QED_A$ at short distances 
as well as the finite proton size. The latter being closely related to the fact that the $s$-states penetrate the nucleus deeply  even for the chosen $g$. This last problem can be relaxed if 
transitions between excited states with nonzero angular momentum are considered instead. Although the problem of their relatively short lifetimes  constitutes a major issue for their experimental 
investigation, their measurements seem to be \emph{a priori} a reliable way to have a  cleaner picture of whether a certain ALP is the cause of the aforementioned discrepancy or not. Inspired by these 
arguments, we consider as an example  the  $3p_{\nicefrac{1}{2}}-3d_{\nicefrac{1}{2}}$ transition. In this case, the required  wave functions are
\begin{equation}
\begin{array}{c}\displaystyle
R_{3p}(r) = \frac{4\sqrt{2}}{9}\frac{1}{(3a_{\mathrm{B}})^{3/2}} \left(  \frac{r}{a_{\mathrm{B}}} \right) \left[1-\frac{r}{6a_{\mathrm{B}}} \right] \mathrm{e}^{-\frac{r}{3a_{\mathrm{B}}}},\\ 
\displaystyle R_{3d}(r) =\frac{2\sqrt{2}}{27\sqrt{5}}\frac{1}{(3a_{\mathrm{B}})^{3/2}} \left(  \frac{r}{a_{\mathrm{B}}} \right)^2  \mathrm{e}^{-\frac{r}{3a_{\mathrm{B}}}}.
\end{array}
\end{equation}An adequate replacement of the radial wave functions in  Eq.~(\ref{lambshift}) by those above allows us to determine the corresponding modification of the transition energy:
\begin{equation}
\begin{split}
&\delta \varepsilon\approx -\frac{\alpha g^2 m^2}{108\pi^2 a_{\mathrm{B}}}\int_1^\infty \frac{du}{u^5}\frac{[u^2-1]^3}{[1+\frac{3}{2}u a_{\mathrm{B}} m]^4}\\&\qquad\qquad\times\left\{1-\frac{2}{1+\frac{3}{2}u a_{\mathrm{B}} m}
+\frac{1}{[1+\frac{3}{2}u a_{\mathrm{B}} m]^2}\right\}.
\end{split}
\label{axionlambshiftfinalfd}
\end{equation}In the limit $a_{\mathrm{B}}\gg\lambda$, the expression above is well approached by $\delta \varepsilon\approx -\alpha g^2/(4374\pi^2m^2 a_{\mathrm{B}}^5)$. The expressions associated with $\rm H_\mu$ 
can be read off from the previous one  by replacing $a_{\mathrm{B}}\to a_{\mathrm{B}}/(186)$.  Taking as a reference the undiscarded region used previously,  $100\ \mathrm{MeV}\lesssim m\lesssim 10\  \mathrm{GeV}$ 
with  $g<10^{-2}\ \rm GeV^{-1}$, Eq.~(\ref{axionlambshiftfinalfd}) has been  evaluated. The result of this assessment is shown in the right panel of Fig.~\ref{fig:mb003}. As in the $2s_{1/2}-2p_{1/2}$ 
transition, the highest energy-shift linked to the $3p_{\nicefrac{1}{2}}-3d_{\nicefrac{1}{2}}$ transition in $\rm H_\mu$ arises from the combination of $g\sim 10^{-2}\ \rm GeV^{-1}$ and $m=10^{-2}\ \rm GeV$. In such a 
case $\delta\varepsilon\sim -10^{-11}\ \rm meV$,  which is five orders of magnitude smaller than the outcome associated with the axion $2s_{1/2}-2p_{1/2}$  Lamb-shift. It is worth emphasizing that, while the 
uncertainty introduced by the  use of spherically symmetric orbitals could be circumvented as described above, the one linked to the physics at distances shorter than $g$ remains. 

%%%%%%%%%%%%%%%%%%%%%%%%%%%%%%%%%%%%%%%%%%%%%%%%%%%%%%%%%%%%%%%%%%%%%%%%%%%%%%%%%%%%%%%%%%%%%%%%%%%%%%%%%%%%%%%%%%%%%%
\subsection{Sensitivity to ALPs  in high-precision  hydrogen spectroscopy}
%%%%%%%%%%%%%%%%%%%%%%%%%%%%%%%%%%%%%%%%%%%%%%%%%%%%%%%%%%%%%%%%%%%%%%%%%%%%%%%%%%%%%%%%%%%%%%%%%%%%%%%%%%%%%%%%%%%%%%

In this section we want to investigate whether the current sensitivity in atomic hydrogen can improve the existing bounds on the axion parameter space. To do this we will analyse the respective
energy-shift in both $2s_{1/2}-2p_{1/2}$ and $1s_{1/2}-2s_{1/2}$ transitions. An expression for the latter $\delta \varepsilon=\int_0^{\infty} dr\, r^2\delta \mathpzc{U}(r)\left[R_{1s}^2(r)-R_{2s}^2(r)\right]$ 
can be easily determined by taking into account the formula for  $R_{2s}(r)$ in Eq.~(\ref{radial2s2p}) and the radial part of the $1s_{\nicefrac{1}{2}}$-state: $R_{1s}(r)=2\exp[-r/a_{\mathrm{B}}]/a_{\mathrm{B}}^{3/2}$. 
Explicitly,
\begin{equation}
\begin{split}
&\delta\varepsilon=
-\frac{\alpha g^2 m^2}{48\pi^2 a_{\mathrm{B}}}\int_1^\infty \frac{du}{u^5}[u^2-1]^3e^{-gm u}\Big\{\frac{1}{[1+\frac{1}{2}u a_{\mathrm{B}} m]^2}\\ &-\frac{1}{2}\frac{1}{[1+u a_{\mathrm{B}} m]^2}
\left(1-\frac{2}{1+u a_{\mathrm{B}} m} +\frac{3}{2[1+u a_{\mathrm{B}} m]^2} \right)\Big\}.
\end{split}
\label{axionlambshiftathydrogen}
\end{equation}

While the left panel in Fig.~\ref{fig:mb004} shows the energy shift for $2s_{1/2}-2p_{1/2}$ [see Eq.~\eqref{axionlambshiftfinal}], the one in the right depicts the result associated with the $1s_{1/2}-2s_{1/2}$ 
transition [see Eq.~(\ref{axionlambshiftathydrogen})]. Both evaluations have been carried out by considering the region of the coupling $10^{-6}\ \mathrm{GeV}^{-1}<g<10^{-2}\ \rm GeV^{-1}$. In contrast to $\mathrm{H}_{\mu}$, 
reliable predictions from high-precision spectroscopy in ordinary hydrogen require to deal with ALPs masses $m \gtrsim 10^{-5}\ \mathrm{GeV}$, corresponding to wavelengths $\lambda\lesssim a_{\mathrm{B}}$. The highest mass shown 
in both panels [$m= 10 \ \rm GeV$] has been set in order to preserve the perturbative condition $mg\ll1$.

When comparing the energy shifts resulting from each panel in Fig.~\ref{fig:mb004} with the corresponding experimental uncertanities [$\vert\delta\varepsilon_{\mathrm{2\sigma}}\vert=2\times 10^{-7}\ \rm meV$ 
for  $2s_{1/2}-2p_{1/2}$ and $\vert \delta \varepsilon_{1\sigma}\vert=1 \times 10^{-6} \ \rm{meV}$ for  $1s_{1/2}-2s_{1/2}$ transition \cite{Roy}] we conclude that, in order to improve the current bounds on the 
axion parameter space, an enhancement in sensitivity of at least five orders of magnitude is required. It is worth remarking that this sensitivity gap also manifests in other high-precision experiments searching 
for potential deviations of the Coulomb's law as those of Cavendish-type. For further details we refer the reader to  Appendix \ref{Appendix}. This lack of sufficient sensitivity  in the context of ALPs is significant 
when taking into account that these setups have allowed for constraining  severely the parameter spaces of other weakly interacting sub-eV particles, including  paraphotons and minicharged particles. However, we should  
emphazise that--in contrast to our investigation--these particle candidates have been treated within renormalizable frameworks and, thus, the bounds have been established on dimensionless coupling constants. 
Likewise, we have already indicated below Eq.~(\ref{counterterms}) that in the axion theory the quantity playing the corresponding role combines two unknown parameters $\sim gm$. Hence, the axion 
mass $m$ suppresses the limits that can be inferred for $g$.  

%%%%%%%%%%%%%%%%%%%%%%%%%%%%%%%%%%%%%%%%%%%%%%%%%%%%%%%%%%%%%%%%%%%%%%%%%%%%%%%%%%%%%%%%%%%%%%%%%%%%%%%%%%%%%%%%%%%%%%%%%%%%%%%%
\section{Conclusion\label{conclus}}
%%%%%%%%%%%%%%%%%%%%%%%%%%%%%%%%%%%%%%%%%%%%%%%%%%%%%%%%%%%%%%%%%%%%%%%%%%%%%%%%%%%%%%%%%%%%%%%%%%%%%%%%%%%%%%%%%%%%%%%%%%%%%%%%%

Within the effective framework of axion quantum electrodynamics, terms  beyond the minimal coupling of two photons to a neutral  pseudoscalar 
field have been used to renormalize the polarization tensor and the axion self-energy operator.  The former outcome was used to establish 
the photon propagator distorted by the quantum vacuum fluctuations of axionlike fields, a piece essential for determining the modification of the Coulomb potential induced by both virtual 
photons and ALPs. This result allowed us to evaluate the way in which atomic spectra could change. Particular attention has been paid to the 
$2s_{\nicefrac{1}{2}}-2p_{\nicefrac{1}{2}}$ transition in hydrogenlike atoms as it might constitute the most natural way of verifying our 
predictions experimentally. Likewise, this sort of axion-modified Lamb-shift has been considered in attempting to explain the proton radius 
anomaly in muonic hydrogen. By contrasting the experimental result with our theoretical prediction, it was found that--up to the uncertainties 
caused by the nature of the transition and the internal limitations of  axion-electrodynamics--ALPs can be excluded  as plausible candidates 
for solving the aforementioned  problem. 

Our investigation has revealed explicitly that neither atomic spectroscopy nor experiments of Cavendish-type allow us to infer bounds that 
improve the existing constraints on the axion parameter space. This fact contrasts with analogous outcomes linked to  scenarios containing minicharged 
particles and hidden photon fields, in which both precision techniques have turned out to be particularly valuable \cite{Popov,Glueck,Jaeckel:2009dh,Roy}. 
The loss of sensitivity within the axion context is conceptually rooted  in the nonrenormalizable character of $\rm QED_A$ and manifests--at the 
level of the modified Coulomb potential Eq.~(\ref{generalpotenatiallwithoutb})--through the dimensionless factor $\sim gm$. This ratio of scales 
accomplishing somewhat a role similar to the coupling strengths of the photon-paraphoton mixing $\chi$ and the parameter $\epsilon$ in the minicharged 
particles scenario. To a certain extent the described problem justifies the existing demand for new laboratory-based routes looking for 
ALPs \cite{evading,Jaeckel:2006id} by using strong electromagnetic fields \cite{Jaeckel:2010ni,Ringwald:2012hr,Hewett:2012ns,Essig:2013lka}, e.g., those 
offered by high-intensity lasers \cite{Di_Piazza_2012,mendonza,Gies:2008wv,Dobrich:2010hi,Villalba-Chavez:2013bda,Villalba-Chavez:2013goa,Villalba-Chavez:2016hxw,Villalba-Chavez:2013gma,Villalba-Chavez:2014nya,Villalba-Chavez:2016hht,Villalba-Chavez:2015xna}.

Let us finally remark that the expression for $\Pi_{\mathrm{R}}^{\mu\nu}(p_1,p_2)$ [see Eq.~(\ref{polarizationtensor})] constitutes an 
essential piece for a more general class of polarization tensors which result  when external electromagnetic fields polarize the vacuum \cite{Villalba-Chavez:2016hxw,VillalbaChavez:2018ea}. 
%Besides, direct applications of this object in the search of ALPs can be envisaged beyond the spectroscopy context. Perhaps the most interesting  
%scenario is the one in which axions drive a tunneling of third kind \cite{Gies:2009wx,Dobrich:2012sw,Gardiner:2012qy,Mou:2017gnm}. 
%Predictions of this quantum effect have been made through minicharged particles and hidden photon fields while the case involving 
%ALPs remains yet undeveloped. An ongoing investigation on this subject is being carried out by the authors.

\begin{acknowledgments}
The authors thank  A.~B.~Voitkiv and A.~E.~Shabad for useful discussions. S. Villalba-Ch\'avez and C. M\"{u}ller gratefully acknowledge  funding by the German 
Research Foundation (DFG) under Grant No. MU 3149/5-1.
\end{acknowledgments}

\appendix

%%%%%%%%%%%%%%%%%%%%%%%%%%%%%%%%%%%%%%%%%%%%%%%%%%%%%%%%%%%%%%%%%%%%%%%%%%%%%%%%%%%%%%%%%%%%%%%%%%%%%%%%%%%%%%%%%%%%%%%%%%%%%%%%%%%%%%%%%%%%%%%%%
\section{Particle-ghost content of the gauge sector and an alternative four-fields formulation of axion-electrodynamics \label{App1}}
%%%%%%%%%%%%%%%%%%%%%%%%%%%%%%%%%%%%%%%%%%%%%%%%%%%%%%%%%%%%%%%%%%%%%%%%%%%%%%%%%%%%%%%%%%%%%%%%%%%%%%%%%%%%%%%%%%%%%%%%%%%%%%%%%%%%%%%%%%%%%%%%%

As mentioned in Sec.~\ref{sec:QKAA}, the photon sector also contains Pauli-Villars ghosts. In order to show this,  let us consider 
the  corresponding Green function resulting from a covariant quantization of  $\bar{a}^\mu(x)$  via a path integral representation. When fixing the gauge via  
$\mathcal{L}_{\rm gauge}=-\frac{1}{2}[(1+\bar{g}^2\bar{\mathpzc{b}}^2_a\square)\partial_\mu \bar{a}^\mu]^2$ it turns out to be  \cite{Bufalo1,Bufalo2}:
\begin{equation}
\begin{split}
&G_{\alpha\beta}(p^2)=\left[-\frac{1}{p^2}+\frac{1}{p^2-\bar{m}_{\rm{gh}}^2}\right]\mathpzc{g}_{\alpha\beta},
\end{split}\label{Podolskypropagator}
\end{equation}where $\bar{m}_{\mathrm{gh}}^2=(\bar{g}\bar{\mathpzc{b}}_{a})^{-2}$ is the corresponding ``bare'' ghost mass. 
Here, a longitudinal  contribution $\sim p_\alpha p_\beta$ has been  ignored on the grounds that, if the photons couple to a conserved current $j^\mu(x)$,  i.e. $p_\mu j^\mu(p)=0$, a term 
of this nature does not contribute to the S-matrix  elements. Manifestly, the photon Green function in  Eq.~(\ref{Podolskypropagator})  resembles Eq.~(\ref{Podolskyscalarpropagator}). However, the particle-ghost 
content linked to this expression is somewhat blurred owing to the presence of the metric tensor $\mathpzc{g}_{\alpha\beta}$. To highlight the emergence of  the Pauli-Villars ghost--leaving 
aside those unphysical  states linked to the quantization procedure that eventually must cancel each other--we  will follow a method that has been  used previously within the context of quantum gravity \cite{VanNieuwenhuizen:1973fi,Sezgin:1979zf,accioly2}.\footnote{We precise that 
higher-derivate operators in combination with nonlocal terms emerge in many other interesting theoretical scenarios, e.g. in construction of effective quantum field theories accounting for quantum conformal and 
chiral anomalies [see e.g. \cite{Giannotti} and references therein].} Rather than  dealing  with the expression above directly, one introduces  the  saturated Green function $\mathscr{G}(p^2)=j^\alpha G_{\alpha\beta}(p^2)j^\beta$ and investigates 
its residues at each pole: $p^2=0$ and $p^2=\bar{m}_{\rm{gh}}^2$. As for any $m\geqslant0$ the relation $j^2\vert_{p^2=m^2}<0$ holds--see proof of Lemma~$1$ in Ref.~\cite{accioly2}--a physical particle is  
linked to a nonnegative  residue of $\mathscr{G}(p^2)$, whereas a ghost emerges  when the contrary occurs. For the case under consideration then follows that  $\left.\mathrm{Res}\;\mathscr{G}(p^2)\right\vert_{p^2=0}>0$ 
(photon) and  $\left.\mathrm{Res}\;\mathscr{G}(p^2)\right\vert_{p^2=\bar{m}_{\rm{gh}}^2}<0$ (ghost). 

Noteworthy, the decompositions of the  axion and photon Green functions [see Eqs.~(\ref{Podolskyscalarpropagator}) and (\ref{Podolskypropagator})] suggest that the  effects of the higher-dimensional 
operators can be formulated in terms of auxiliary--fictitious--fields.  In this context, the  action of interest reads
\begin{equation}
\begin{split}\label{prac4field}
&\mathcal{S}_{\mathrm{eff}}=\int d^4x\;\left\{-\frac{1}{4}f_{\mu\nu}f^{\mu\nu}+\frac{1}{2}\partial_\mu\bar{\phi}\partial^\mu\bar{\phi}-\frac{1}{2}\bar{m}^2\bar{\phi}^2\right.\\
&\qquad\quad -\Phi\square\bar{\phi}+\frac{1}{2}\bar{m}_{\mathrm{s}}^2\Phi^2-\frac{1}{2} \bar{m}^2_{\mathrm{gh}}\mathpzc{A}^2+\frac{1}{2}f_{\mu\nu}\mathpzc{F}^{\mu\nu}\\
&\qquad\quad+\left.\frac{1}{4}\bar{g}\bar{\phi}\tilde{f}_{\mu\nu}f^{\mu\nu}\right\},
\end{split}
\end{equation}where $\mathpzc{F}_{\mu\nu}=\partial_\mu \mathpzc{A}_\nu-\partial_\nu \mathpzc{A}_\mu$ and  $\bar{m}_{\mathrm{s}}^2=(\bar{g}\bar{\mathpzc{b}}_{\phi})^{-2}$. 
We remark that  the equations of motion for the auxiliary fields are exact 
\begin{equation}
\Phi=\frac{1}{\bar{m}_{\mathrm{s}}^2}\square\bar{\phi},\quad \mathpzc{A}_{\lambda} =-\frac{1}{\bar{m}_{\mathrm{gh}}^2}\partial_\mu f^{\mu}_{\ \lambda}. 
\end{equation}Hence, when integrating out  both $\Phi(x)$ and  $\mathpzc{A}_{\lambda}(x)$ classically,  i.e. by removing them from $\mathcal{S}_{\mathrm{eff}}$ using their equations of motion, we 
reproduce the action of axion-electrodynamics extended by terms proportional to  $\bar{g}^2$, i.e. $\mathcal{S}_{\mathrm{eff}}\to \mathcal{S}=S_{\bar{g}}+S_{\bar{g}^2}$ [see Eqs.~(\ref{initialaction}) and (\ref{PLWaction})].

Observe that, as a consequence of the shift $\bar{a}\to a+\mathpzc{A}$ and $\bar{\phi}\to \phi-\Phi$, the functional action in  Eq.~(\ref{prac4field}) can be written as 
\begin{equation}
\begin{split}
&\mathcal{S}_{\mathrm{eff}}=\int d^4x\;\left\{-\frac{1}{4}f_{\mu\nu}f^{\mu\nu}+\frac{1}{4}\mathpzc{F}_{\mu\nu}\mathpzc{F}^{\mu\nu}-\frac{1}{2} \bar{m}^2_{\mathrm{gh}}\mathpzc{A}^2\right.\\
&\qquad\quad+\frac{1}{2}\partial_\mu\phi\partial^\mu\phi-\frac{1}{2}\bar{m}^2(\phi-\Phi)^2-\frac{1}{2}\partial_\mu\Phi \partial^\mu\Phi\\
&\qquad\quad+\frac{1}{2}\bar{m}_{\mathrm{s}}^2\Phi^2+\frac{1}{4}\bar{g}(\phi-\Phi)\left[\tilde{f}_{\mu\nu}f^{\mu\nu}+2\tilde{\mathpzc{F}}_{\mu\nu}f^{\mu\nu}\right.\\
&\qquad\quad\left.\left.+\tilde{\mathpzc{F}}_{\mu\nu}\mathpzc{F}^{\mu\nu}\right]\right\}.\nonumber
\end{split}
\end{equation}Clearly, the  first term contained in this formula  is the Maxwell Lagrangian, while the combination of the two remaining contributions in the first line somewhat looks like  
the Proca Lagrangian,  with the  exception that the sign of its  kinetic term is not the usual one. Likewise, the kinetic portion linked to  the $\Phi(x)$ field manifests an opposite  sign 
to the corresponding contribution of $\phi(x)$. Owing to the described feature the associated Hamiltonian is not positive defined leading--upon  quantization--to the absence of a ground 
state. The particle-ghost content of the theory  is elucidated in this alternative formulation to the requirement that free fields of 
particles (ghosts) have positive (negative) energy. 

%%%%%%%%%%%%%%%%%%%%%%%%%%%%%%%%%%%%%%%%%%%%%%%%%%%%%%%%%%%%%%%%%%%%%%%%%%%%%%%%%%%%%%%%%%%%%%%%%%%%%%%%%%%%%%%%%%%%%%%%%%%%%%%%%%%%%%%%%%%%%%%%%
\section{Sensitivity to ALPs in experiments of Cavendish-type \label{Appendix}}
%%%%%%%%%%%%%%%%%%%%%%%%%%%%%%%%%%%%%%%%%%%%%%%%%%%%%%%%%%%%%%%%%%%%%%%%%%%%%%%%%%%%%%%%%%%%%%%%%%%%%%%%%%%%%%%%%%%%%%%%%%%%%%%%%%%%%%%%%%%%%%%%%

Precision tests of   Coulomb's law via Cavendish-type experiments have severely constricted  the parameter space of hidden photons in the $\mu\rm eV$  mass regime \cite{Okun:1982xi,Popov}.
Today, these setups also  provide  the best laboratory bounds on mini-charged particles in the sub$-\mu\rm eV$ range \cite{Jaeckel:2009dh}. Here, we want to estimate the sensitivity 
of this type of experiments in the context of ALPs. To this end, we consider a setup containing two concentric spheres: an outer charged conducting sphere--characterized by a radius  
$b$--and an uncharged conducting inner sphere with a radius $a$. Only if the electrostatic potential follows the $r^{-1}$ law, the potential difference between the spheres vanishes 
and the cavity is free of electromagnetic field. However, deviations from the Coulomb potential like those induced by loop corrections [compare Eq.~\eqref{generalpotenatiallwithoutb}] 
could lead to a nontrivial relative voltage difference $\gamma_{ab}=\vert [\mathpzc{u}(\mathpzc{Q},b,b)-\mathpzc{u}(\mathpzc{Q}, a,b)]/\mathpzc{u}(\mathpzc{Q},b,b)\vert$ that can be detected. This observable depends 
on  the potential of the charged sphere evaluated on its surface $\mathpzc{u}(\mathpzc{Q},b,b)$ as well as on the surface of the inner sphere $\mathpzc{u}(\mathpzc{Q},a,b)$.

In general, the electrostatic potential of a sphere with  radius $b$ and  charge $\mathpzc{Q}$ evaluated at a distance $r$ from its center has the form
\begin{equation}\label{A1}
\begin{array}{c}\displaystyle
\mathpzc{u}(\mathpzc{Q},r,b) = \frac{\mathpzc{Q}}{2br}[f(r+b) -f(|r-b|)],\\ \\
f(r) = \int_0^r s \mathpzc{A}_0(s) ds,
\end{array}
\end{equation}where $\mathpzc{A}_0(s)$ is an arbitrary potential in which the charge of the pointlike particle must be set to unity \cite{Maxwell}. Now, to determine $\gamma_{ab}$ resulting from $\mathrm{QED_A}$ 
we insert the axion-modified Coulomb potential $\mathpzc{a}_0(s,\mathpzc{q} = 1)$ [see Eq.~\eqref{generalpotenatiallwithoutb}] into the expression above. Notice that, similarly to the case analyzed in Sec.~\ref{ACPsecb}, 
the integral over $s$ must be split into two parts: $ \int_0^{r} ds\ldots= \int_0^{g} ds\ldots+ \int_{ g}^{r} ds\ldots$. Ignoring the contribution coming from the region $[0,g]$ we obtain
{\begin{equation}
\begin{split}
&\gamma_{ab} \approx \Big|
\frac{ g^2 m}{96\pi^2}\int_1^\infty \frac{du}{u^6}[u^2-1]^3\left\{ \frac{e^{-2bm u}-e^{-gm u}}{b} \right.\\ 
&\qquad \left.+ \frac{2}{a} e^{-bm u}\mathrm{sinh}(amu)\right\} + \mathcal{O}(g^4m^2)\Big|.
\end{split}\label{A2}
\end{equation}}The integral that remains in this formula can be calculated analytically by using (3.351.4) in Ref.~\cite{Gradshteyn}. Since both $b$ and $a$ are macroscopic quantities, the conditions $a,b, b-a \gg g$ hold 
and we can approximate the expression above by
{\begin{equation}\label{gammaab}
\begin{split}
&\gamma_{ab} \approx \Big|
\frac{ g^2 m}{96\pi^2 b }\int_1^\infty \frac{du}{u^6}[u^2-1]^3 e^{-gm u}\Big|.
\end{split}
\end{equation}}Notice that Eq.~\eqref{gammaab} is independent of the radius $a$ of the inner sphere. When considering the limit $mg\ll1$ we obtain
\begin{equation}\label{gammaab1}
\gamma_{ab} \approx 
\frac{g}{96\pi^2 b},
\end{equation} which does not depend on the axion mass $m$ either.

With $\gamma_{ab}$ to our disposal, we can proceed to estimate the sensitivity of this setup in the search for ALPs. For such a purpose, we use the benchmark parameters of the experiment performed by 
Plimpton and Lawton [$a=38\ \rm cm$, $b=46\ \rm cm$] in which a margin for $\gamma_{ab}$ exists,  provided it lies below $\vert\gamma_{ab}\vert\lesssim 3\times 10^{-10}$ \cite{Plimpton}. By making use of Eq.~\eqref{gammaab1} 
we find 
\begin{equation}\label{cavedishconstraint}
g  \lesssim 6.7 \times 10^{7} \ \mathrm{GeV}^{-1}.
\end{equation}We emphasize that--as a consequence of the perturbative condition [see below Eq.~(\ref{gammaab})]--this result applies for $m \ll 15 \ \mathrm{eV}$. Besides,  it  is trustworthy for axion 
wavelengths smaller than the typical length scale of the spheres $\sim 0.1\ \rm m$, i.e., for axion  masses $m \gg 10^{-6} \ \mathrm{eV}$. We note that the constraint in Eq.~(\ref{cavedishconstraint}) for the 
mass region $10^{-6}\ \mathrm{eV}\ll m\ll15\  \rm eV$ has already been  discarded by combining the experimental outcomes of collaborations such as PVLAS and OSCAR  [see Fig.~\ref{fig:mb005}]. 
Hence, the sensitivity in this experiment of Cavendish-type is not high enough to improve the existing bounds on the axion parameter space. 

It is worth remarking that a more accurate version of this kind of experiments has been carried out by  using four concentric icosahedrons \cite{Williams}. For obtaining first estimates, they may be treated 
approximately as four concentric spheres. In contrast to the setup of Plimpton and Lawton, here a very high voltage is applied between the outer two spheres with radii $d=127\ \rm cm$ and $c=94.7\ \rm cm$. 
The voltage difference is measured between the two internal ones, with radii $b=94\ \rm cm$ and $a=60 \ \rm cm$, which are uncharged. This setup allows us to  infer bounds for the ALPs parameters via the ratio 
between the voltage differences: 
\begin{equation}\label{fourisoca}
\gamma_{abcd}=\Big|\frac{\mathpzc{u}(\mathpzc{Q},c,b)-\mathpzc{u}(\mathpzc{Q},d,b) -\mathpzc{u}(\mathpzc{Q},c,a) + \mathpzc{u}(\mathpzc{Q},d,a)}{2\mathpzc{u}(\mathpzc{Q},c,d)-\mathpzc{u}(\mathpzc{Q},d,d) -\mathpzc{u}(\mathpzc{Q},c,c)}\Big|.
\end{equation} We insert Eq.~\eqref{generalpotenatiallwithoutb} into \eqref{A1} and evaluate the resulting formula in the various parameters contained in Eq.~(\ref{fourisoca}). As a consequence, we end up with 
\begin{equation}
\begin{split}
&\gamma_{abcd}= \Big|
\frac{g^2m c}{48\pi^2\delta}\int_1^\infty du\left[1-\frac{1}{u^2}\right]^3 e^{-md u} \\ &\quad\times\left[1 -\frac{d}{c}e^{-m\delta u}\right]\left[\frac{\mathrm{sinh}(mbu)}{b} - \frac{\mathrm{sinh}(mau)}{a} \right]\Big|,
\end{split}\label{A7}
\end{equation}where $\delta\equiv c-d$ and  terms of the order of $\sim g^4m^2$ have been disregarded. Notice that, in contrast to Eq.~(\ref{A2}), the integrand above lacks  terms involving $\sim e^{-mg u}$. 
This could be anticipated because the numerator of  $\gamma_{abcd}$ [see Eq.~(\ref{fourisoca})]  does not contain a potential evaluated at the surface of the  spheres [compare with $\gamma_{ab}$ given above 
Eq.~(\ref{A1})]. Consequently,  when the condition $md=d/\lambda\ll 1$ is satisfied, the asymptotic expression for $\gamma_{abcd}$ becomes 
independent of the axion mass and quadratic in $g$:
\begin{equation}
\begin{split}\label{finaldredococ}
& \gamma_{abcd}  \approx \left\vert\frac{ g^2}{96\pi^2}\frac{c}{b(c-d)(d-b)}\right. \\
&\qquad \left.\times\left(1 -\frac{b(d-b)}{a(d-a)}-\frac{d(d-b)}{c(c-b)} + \frac{db(d-b)}{ac(c-a)}\right)\right\vert.
\end{split}
\end{equation}Since  this formula  applies for axion wavelengths larger than the typical length scale of the experiment,  the outcomes resulting from it  can be considered 
reliable as long as the interactions between ALPs and plausible fields/matter existing outside of the  external icosahedron are negligible.  Next, the aforementioned experiment achieves a precision 
$\vert \gamma_{abcd}\vert\lesssim2\times 10^{-16}$ \cite{Williams,Jaeckel:2009dh}. Combining this value  with Eq.~(\ref{finaldredococ}) 
we  constraint $g$ to lie below
\begin{equation}
g  \lesssim 9.8 \times 10^{7} \ \mathrm{GeV}^{-1} \quad \mathrm{for}\quad  m \ll 10^{-7} \ \mathrm{eV}.
\end{equation}Noteworthy, despite  the improvement in the experimental accuracy, the resulting upper limit turns out to be comparable to the one found from the results of Plimpton and Lawton 
[see Eq.~(\ref{cavedishconstraint})] and so, no  improvement is found as compared with the existing constraints.  The lack of sensitivity is understood here as a direct consequence of the quadratic 
dependence of $\gamma_{abcd}$ on  $g$  [see Eq.~(\ref{finaldredococ})].

\end{document}